\documentclass{aa}
\usepackage{natbib}
\usepackage{txfonts}
\usepackage{graphicx}
\begin{document}

\title{The spin of the second-born black hole in coalescing binary black holes}

\author{Y. Qin\inst{1,\dag}, T. Fragos \inst{2,1}, G. Meynet\inst{1}, J. Andrews\inst{3,4}, M. S\o rensen\inst{1}, H.F. Song\inst{1,5,6}}
\institute{Geneva Observatory, University of Geneva, CH-1290 Sauverny, Switzerland
\and DARK, Niels Bohr Institute, University of Copenhagen, 2100 Copenhagen, Denmark
\and Foundation for Research and Technology - Hellas, IESL, Voutes, 71110 Heraklion, Greece
\and Physics Department \& Institute of Theoretical \& Computational Physics, University of Crete, 71003 Heraklion, Crete, Greece
\and College of Physics, Guizhou University, Guiyang City, Guizhou Province, 550025, P.R. China
\and Key Laboratory for the Structure and Evolution of Celestial Objects, Chinese Academy of Sciences, Kunming 650011
}
\offprints{Ying Qin, \email{ying.qin@unige.ch}}

\titlerunning{The spin of the second-born BH in coalescing BBHs}
\authorrunning{Qin et al.}

\abstract
{Various binary black hole formation channels have been proposed since the first gravitational event GW150914 was discovered by the Advanced Laser Interferometer Gravitational-Wave Observatory (AdLIGO). For all evolutionary channels based on the evolution of isolated binaries, the immediate progenitor of the binary black hole is a close binary system composed of a black hole and a helium star.}
{We study the spin angular momentum evolution of the helium star in order to constrain the spin of the second-born black hole.}
{We perform detailed stellar structure and binary evolution calculations that take into account, mass-loss, internal differential rotation, and tidal interactions between the helium star and the black hole companion, where we also calculate the strength of the tidal interactions from first principles based on the structure of the helium stars.
We systematically explore the parameter space of initial binary properties, including initial black hole and helium star masses, initial rotation of the helium star as well as metallicity.}
{We argue that the spin of the first-born black hole at its birth is negligible ($\lesssim 0.1$), hence the second-born black hole's spin dominates the measured effective spin, $\chi_{\rm eff}$, from gravitational wave events of double black hole mergers. We find that tides can be important only when orbital periods are shorter than 2 days. Upon core collapse, the helium star produces a black hole (the second-born black hole in the system) with a spin that can span the entire range from zero to maximally spinning. We show that the bimodal distribution of the spin of the second-born black hole obtained in recent papers is mainly due to oversimplifying assumptions. We find an anti-correlation between the merging timescale of the two black holes, T$_{\rm merger}$, and the effective spin $\chi_{\rm eff}$. Finally, we provide new prescriptions for the tidal coefficient E$_2$ for both H-rich and the helium-rich stars.}
{To understand the spin of the second-born black hole, careful treatments of both tides and stellar winds are needed. We predict that, with future improvements to AdLIGO's sensitivity, the sample of merging binary black hole systems will show an overdensity of sources with positive but small $\chi_{\rm eff}$ originating from lower mass black hole mergers born at low redshift.}

\keywords{ binaries: close binary stars; gravitational waves; stars: Wolf-Rayet; black hole}

\maketitle

\section{Introduction}

Stellar-mass black holes (BHs) are formed from the gravitational collapse of massive stars \citep[$\gtrsim$ 20 M$_\odot$; e.g.][]{1999ApJ...522..413F,2003ApJ...591..288H,2016ApJ...821...38S} after they  exhaust the nuclear fuel at their centers. Astrophysical BHs can be fully described by only two quantities; their mass, M, and angular momentum $\vec{J}$. The angular momentum content of a BH is usually described by  the dimensionless BH spin parameter
\begin{equation}
\vec{a}  = c\vec{J}/GM^2  \text{ , }
\end{equation}
where $c$ is the speed of light in vacuum. Many BHs exist in binary systems with non-degenerate companion stars, e.g. X-ray binaries, which makes it possible to obtain the BH's properties indirectly \citep{2006AAS...208.3301M,2014SSRv..183..295M,2014SSRv..183..277R,2014SSRv..183..223C,2015PhR...548....1M}. Binary systems where both members are BHs can also exist. If the orbital separation of these binary BHs (BBHs) is initially sufficiently small, angular momentum losses due to gravitational wave emission contract their orbit further and can lead to their coalescence within the Hubble time. The existence of such BBHs in the Universe, as a result of the evolution of massive binary stars, was first theorized by \citet{1973NInfo..27...70T}.

A new window for the study of the BHs has opened with the detection of the first gravitational wave event \citep{2016PhRvL.116f1102A} by the Advanced Laser Interferometer Gravitational-Wave Observatory (AdLIGO) \citep{2015CQGra..32g4001L}. To date, six gravitational wave events and one high-significance gravitational wave event candidate \citep{2016PhRvX...6d1015A,2016PhRvL.116x1103A,2016PhRvL.116f1102A,2017PhRvL.118v1101A,2017PhRvL.119p1101A,2017ApJ...851L..35A} from the merger of BBHs have been detected by AdLIGO. These observations demonstrate that massive stellar BBHs exist and can merge within the Hubble time \citep{2016ApJ...818L..22A}.
The suggested formation channels of these BBHs can be split into two broad categories: (i) the formation via the evolution of massive, isolated binaries in the field, that either go through a common envelope (CE) phase after the formation of the first-born BH that significantly shrinks their orbits, hereafter referred as the ``CE'' binary evolution channel \citep{1991ApJ...380L..17P,1993MNRAS.260..675T,2016Natur.534..512B}, or possibly a stable, non-conservative mass-transfer phase \citep{2017MNRAS.471.4256V, 2017MNRAS.468.5020I}, or spend their whole lives in close orbits and evolve chemically homogeneously \citep{2016MNRAS.460.3545D,2016A&A...588A..50M,2016MNRAS.458.2634M,2016A&A...585A.120S}, and (ii) the dynamical formation in globular cluster \citep{1993Natur.364..423S,2015PhRvL.115e1101R,2016ApJ...824L...8R} and galactic nuclear clusters \citep{2009ApJ...692..917M,2009MNRAS.395.2127O,2012PhRvD..85l3005K,2017ApJ...846..146P}.
Finally, motivated by the existence of a potential electromagnetic counterpart for GW150914 in Gamma-rays \citep{2016ApJ...826L...6C}, a formation channel from a single star, via the fragmentation of their rapidly rotating cores, has been suggested \citep{2016ApJ...819L..21L,2017arXiv170604211D}. However, both the applicability of this scenario to GW150914 \citep{2016ApJ...824L..10W} and the detection of the electromagnetic counterpart itself \citep{2016ApJ...820L..36S,2016ApJ...823L...2A} have been questioned.

For the isolated field binary channels, the spins of the two BHs are expected to be mostly aligned  with the orbital angular momentum. In contrast, the spins of the two BHs from the dynamical formation channels are expected to have a random, isotropically distributed direction. Clearly the spin plays an important role in distinguishing among the various BBH formation channels \citep[e.g.][]{2016ApJ...818L..22A,2017Natur.548..426F}. The spins of the BHs have an effect on the waveform of the gravitational waves, and this can be observed by AdLIGO. The effective inspiral spin parameter $\chi_{\rm eff}$ which can be directly constrained by the gravitational wave signal, is defined in the following expression:
\begin{equation}\label{xeff}
    \chi_{\rm eff} \equiv \frac{M_1\vec{a_1}+M_2\vec{a_2}}{M_1+M_2}\vec{\hat{L}},
\end{equation}
where M$_1$ and M$_2$ are the masses of the two BHs, $\vec{a_1}$ and $\vec{a_2}$ are two BHs' dimensionless spin parameters and $\vec{\hat{L}}$ is the direction of the orbital angular momentum. $\chi_{\rm eff}$ has been observed to be $- 0.06_{-0.14}^{+0.14}$, $0.21_{-0.1}^{+0.2}$, $-0.12^{+21}_{-30}$, $0.0_{-0.2}^{+0.3}$, $0.06_{-0.12}^{+0.12}$ and $0.07_{-0.09}^{+0.23}$, for GW150914, GW151226, GW170104, GW170814,  LVT151012 and GW170608, respectively \citep{2016PhRvX...6d1015A,2016PhRvL.116x1103A,2017PhRvL.118v1101A,2017PhRvL.119n1101A,2017ApJ...851L..35A}. From these 6 $\chi_{\rm eff}$ measurements, 5 are consistent with 0 and only for GW151226 $\chi_{\rm eff}$ is determined to have a positive, non-zero value with a high statistical confidence. Assuming an isotropic prior probability distribution for the misalignment angle between the individual BH spins and the orbit, the individual BH spins of GW170104 have a significant probability of being misaligned with the orbit, supporting the dynamical formation scenario. Alternatively, if the individual BH spin magnitudes are small, then the posterior probability of a misalignment between the individual BH spins with the orbit decreases and the ``CE'' channel can not be ruled out \citep{2017arXiv170607053B}.

In all the formation channels that are based on the evolution of an isolated field binary (i.e. the ``CE'' binary evolution channels and the chemically homogeneous evolution channels), the immediate progenitor of the BBH is a close binary consisting of a BH and a helium (He)-rich star (i.e. WR star). In these binaries, the angular momentum of the progenitor of the second-born BH will be mainly determined by the net effect of the stellar winds and the tidal interaction in a close binary configuration. On the one hand, the outer layer of the He-rich star will be lost through stellar winds with a mass loss rate strongly dependent on the metallicity of the mass-losing He-rich star \citep[e.g.][]{2006A&A...452..295E}. At the same time, this mass loss rate can be potentially enhanced if the star is rapidly rotating, approaching critical rotation \citep{1997ASPC..120...83L,2000A&A...361..159M}. As a result of the mass and  rotational and orbital angular momentum losses due to stellar winds, both the orbital separation and the rotation period of the He-rich star change. Stellar winds tend to increase the rotation period of the mass-losing star. As a result of the mass and angular momentum losses due to stellar winds, both the orbital separation and the rotation period of the He-rich star tend to increase; stellar winds extract both spin angular momentum from the mass-losing star, slowing its rotation, and mass and orbital angular momentum from the system, tending to widen the binary orbit. In addition to these effects, tidal interactions between the BH and the He-rich star may also induce angular momentum exchanges between the orbit and the star.
Apart from stellar winds and tides, different initial conditions for the He-rich star at its birth, including initial rotation rate and metallicity, also play a key role in the spin of the second-born BH at its birth.

Following the detection of the first gravitational wave event, GW150914, for which the quantity $\chi_{\rm eff}$ of the two coalescing BHs was found to be consistent with zero, several studies attempted to model this last evolutionary phase in the formation of a BBH and derive constraints on the expected spin of the second-born BH, under the assumption that these BBH were formed via the ``CE'' channel \citep{2016MNRAS.462..844K,2017arXiv170708978H,2018MNRAS.473.4174Z}. These studies employed analytic arguments and semi-analytic calculations to infer the angular momentum content of the progenitor of the second-born BH due to tidal interactions with its BH companion. The main conclusion of these studies was that the  distribution of the second-born BH is expected to be bi-modal, with approximately half having no spin and half spinning maximally. However, in order to make the problem analytically tractable, in all three studies, several simplifying assumptions had to be made. For example, they used approximate timescales for the process of tidal synchronization that do not take into account changes in the structure of the star during its lifetime. Furthermore, these studies did not self-consistently take into account wind mass-loss which, through tidal coupling, affects the evolution of the orbit and the angular momentum content of the WR star. Hence, it is not obvious that these results will persist when using detailed binary evolution models that self-consistently include the complex interplay between tides, wind mass loss and stellar structure evolution.

Massive He-rich stars are also widely accepted as the progenitors of type Ibc supernovae \citep[e.g.][and references therein]{1997ARA&A..35..309F,2006ARA&A..44..507W}. During the core collapse of a rapidly rotating He-rich star, its outer layers may form an accretion disk along with a highly relativistic jet around the newly formed BH resulting in the release of intense Gamma-ray radiation. According to this paradigm, also known as the collapsar model \citep{1993AAS...182.5505W}, massive, rapidly rotating He-rich stars are the progenitors of long Gamma-Ray Bursts (lGRBs). In this context,  the tidal spin-up of a He-rich star from a compact companion in a close binary system has been investigated in several studies \citep{2000NewA....5..191B,2004MNRAS.348.1215I,2005A&A...435.1013P,2007A&A...465L..29C,2007Ap&SS.311..177V,2008A&A...484..831D,2008MNRAS.384.1109E}. Angular momentum can be transferred through the L-S coupling effect from the orbit to the He-rich star, but the orbital period needs to be sufficiently short to allow for a strong tidal interaction. Transferred angular momentum from the orbit to the outer layers of the He-rich star will spin up its core by various coupling effects such as toroidal magnetic fields generated from differentially rotating, radiative stellar envelopes \citep{1999A&A...349..189S,2002A&A...381..923S}.

In this paper, we focus on the later phases of the ``CE'' BBH formation channel and specifically on the evolution of a close binary system consisting of a He-rich star and a BH, immediately after the binary detaches at the end of the CE phase. Our results are relevant for all BBH formation channels based on the evolution of an isolated field binary. We investigate the angular momentum content of the second-born BH progenitor, using detailed stellar structure and binary evolution calculations that take into account the effects of internal differential rotation in the He-rich star, stellar winds, and tides. In order to better understand the interplay of these effects, we explore a 5-dimensional initial parameter space of initial masses of the two binary components, initial orbital period, initial rotation of the He-rich star and metallicity.

This paper is organized as follows. In \S2, we present qualitative arguments about the expected spin of the first-born BH in the ``CE'' isolated binary evolution channel. We then introduce the theory of tidal interaction adopted in this study in \S3. In \S4, we show a semi-analytic test for the efficiency of tides in WR-BH binaries. In \S5, we present detailed simulations of the angular momentum evolution of He-rich stars in close binary systems. In \S6 we discuss the merging timescale of the two compact objects. Finally, discussion and conclusions of our results are given in \S7 and \S8, respectively.

\section{The spin of the first-born BH in the classical isolated binary evolution channel}

In the ``CE'', isolated field binary formation channel \citep[for instance, see][]{2016Natur.534..512B}, a binary consisting of two massive hydrogen (H)-rich stars in a wide orbit evolve from Zero Age Main Sequence (ZAMS). The more massive star (star 1) evolves faster and fills its Roche lobe during either the Hertzsprung gap or the supergiant phase. Star 1 transfers mass to the less massive star (star 2) through the first Lagrangian point, and the mass transfer (MT) during this phase is stable. After losing all its H envelope, star 1 evolves into a He star and soon directly collapses to form a BH (the first-born BH), while star 2, which has accreted part of star 1's envelope, still remains on the MS. At this point, the orbital separation has increased further, mainly due to MT. Subsequently, star 2 evolves off the main sequence and overfills its Roche lobe while on the red supergiant branch. Because of the mass ratio and the evolutionary stage of star 2, this MT phase is dynamically unstable and the binary enters into a CE phase. The BH spirals into the envelope of star 2, converting orbital energy into heat. During this phase the orbital separation shrinks dramatically, and the post-CE system consists of a He star and a BH in a close orbit of tens of Solar radii. Eventually star 2 also collapses to form a BH, and potential asymmetries in the core collapse may alter the orbit further. The final product of this formation channel is a BBH in a close enough orbit that can lead to the coalescence of the two BHs due to angular momentum losses from gravitational wave emission.

In this scenario, it is expected that the spin of the first-born BH is very low ($a_{*,1}\sim 0$) for two reasons. First, while the progenitor of star 1 evolves through a red supergiant phase  \citep[assuming an efficient angular momentum transport mechanism such as the Taylor-Spruit dynamo;][]{1999A&A...349..189S,2002A&A...381..923S}, most of the initial angular momentum is transported to the outer layers of the star upon expansion. The core of the star, although still rotating at a higher angular frequency than the envelope, is depleted of angular momentum. Eventually, the outer layer of the red supergiant star is removed either by the MT phase or by stellar winds, and thus the remaining angular momentum in the core of the star will be small. Second, before the onset of the MT phase, the orbital separation is quite large. Thus, even if tides can efficiently synchronize the rotation of the outer layers of the star to the orbit, the angular frequency of the latter will be so low that it will not be possible to actually spin up the core. A similar argument has been presented by \citet{2015ApJ...800...17F} about the natal spin of BHs in Galactic low-mass X-ray binaries.

To obtain a more quantitative handling on the arguments presented above, we evolved, using the Modules for Experiments in Stellar Astrophysics (MESA) code version 8118 \citep{2011ApJS..192....3P,2013ApJS..208....4P,2015ApJS..220...15P,2018ApJS..234...34P}, single massive stars of 50 M$_\odot$ and 90 M$_\odot$ at metallicities of 0.01 Z$_\odot$, 0.1 Z$_\odot$ and Z$_\odot$ (Z$_\odot$ is the solar metallicity taken here as 0.02). For each mass and metallicity, we evolve the stars from ZAMS, assuming different initial rotation rates (i.e. 0.1, 0.3, 0.5, 0.7 and 0.9 $\omega_{\rm init}/\omega_{\rm crit}$; where $\omega_{\rm init}$ and $\omega_{\rm crit}$ are the initial and the critical angular velocity at the surface of the star), where we assume that the stars, initially, are uniformly rotating. For this set of single H-rich models, we use stellar winds, mixing and angular momentum transport parameters closely following \citet{2016A&A...588A..50M}. We also use the Schwarzschild criterion to treat the boundary of the convective zones and a convective core overshooting parametrized with $\alpha_{ov}= 0.1$. The impact of rotation on the mass loss rate is considered as indicated in Eq. \ref{ml} \citep{1998A&A...334..210H,1998A&A...329..551L}.
    \begin{equation}\label{ml}
    \centering
    \dot{M}(\omega)= \dot{M}(0)\left(\frac{1}{1-\omega/\omega_{\rm crit}}\right)^\xi,
    \end{equation}
where $\omega$ and $\omega_{\rm crit}$ ($\omega_{\rm crit}^2 = (1- L/L_{\rm Edd})GM/R^3$, $L_{\rm EDD}$ is the Eddington luminosity) are the angular velocity and critical angular velocity at the surface, respectively. The default value of the exponent $\xi = 0.43$ is taken from \citet{1998A&A...329..551L}. No gravity darkening effect is accounted for \citep[see][for a discussion on the impact of this process]{2000A&A...361..159M}.
More details about our settings in MESA for the single H-rich stars can be found on the MESA web page \footnote{The detailed list of parameters used for the single H-rich stars can be found at http://mesastar.org/results.}.

\begin{figure}[h]
     \centering
     \includegraphics[width=\columnwidth]{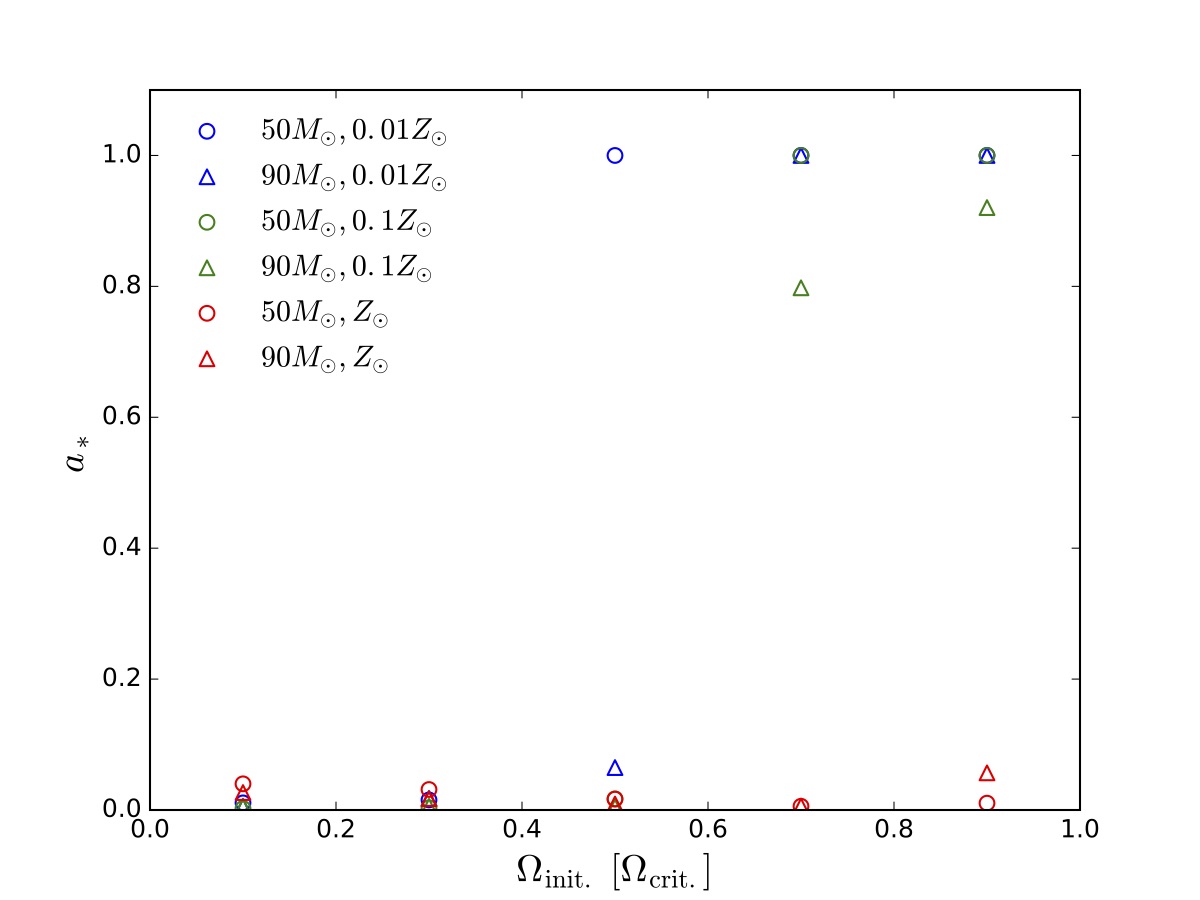}
     \caption{ The spin of the resultant BH from the evolution of single stars at central He exhaustion for different masses (50 M$_\odot$ and 90 M$_\odot$) and metallicities (0.01 Z$_\odot$, 0.1 Z$_\odot$, and Z$_\odot$) as a function of initial rotation, assuming that the carbon/oxygen core will collapse to form the BH, conserving the angular momentum it had at the point of central He exhaustion.}
     \label{fig1}
\end{figure}

We run all the models up to central He exhaustion. The spin of the BH is obtained
assuming that its mass and angular momentum content are given by the mass and angular momentum of the carbon-oxygen core at that stage. The final spin obtained as indicated above is shown in Fig.~\ref{fig1}
as the function of the initial relative rotation rate at ZAMS.
We find that for initial rotations up to 0.5 $\omega_{\rm init}/\omega_{\rm crit}$ and for all metallicities, the spin of the resultant BH is negligible ($\lesssim 0.1$).
Although Fig.~\ref{fig1} shows that stars with initial relative rotations above 0.5 $\omega_{\rm init}/\omega_{\rm crit}$ and metallicities $\lesssim$0.1 Z$_\odot$ produce near maximally spinning BHs, the fast-spinning nature of these He-stars progenitors induces efficient internal mixing, forcing chemically homogeneous evolution; these stars never evolve onto the supergiant branch, and hence they cannot be progenitors of the first-born BH in the ``CE'' formation scenario we consider here.
Overall, we expect, as a first order approximation, that the first-born BH in the ``CE'', isolated field binary formation channel has negligible spin.

The latter argument is even more true when accounting for two effects that we have neglected in this approach and that would remove angular momentum from the star. First, in this approach
we considered that the star evolves as a single star, while it is a member of a binary system. As recalled above, in a close binary, the more massive star loses angular momentum through the first Lagrangian point during the stable MT phase, and its angular momentum content decreases. Second, we have neglected the effects of disk formation during the core collapse process \citep[e.g. see discussion in section 2.2 of][]{2017arXiv170607053B} which may remove some angular momentum from the core that collapses to form the BH. The latter effect is also relevant for the estimate of the spin of the second-born BH that we discuss later. In that sense our quoted spins can be considered as upper limits for the predicted BH spin.

\section{Tidal interaction in binary systems}

Tidal forces, in a close binary system, play a key role in both secular evolution of the orbit and the internal angular momentum of the stellar components. Two main mechanisms responsible for the dissipation of the tidal kinetic energy have been widely accepted, i.e., turbulent dissipation (or convective damping) on the equilibrium tides applied to the stars with an outer convection zone, and radiative damping on the dynamical tides applied to the stars with an outer radiative zone \citep{1977A&A....57..383Z}. The strength of the interaction depends on the ratio of the stellar radius to the orbital separation of the two stars, and the timescale of synchronization is defined  as follows \citep{1977A&A....57..383Z,1981A&A....99..126H}:
    \begin{equation}\label{tsync}
    \centering
    \frac{1}{T_{\rm sync}} = -\frac{1}{\omega-n} \frac{\mathrm{d}\omega }{\mathrm{d} t} = 3\cdot \frac{K}{T}\frac{q^2}{r^2_{g}}\left(\frac{R}{a}\right)^6,
    \end{equation}
where $\omega$ and $n$ are the spin angular velocity and the orbital angular velocity, respectively, $q$ is the mass ratio of the secondary star to the primary one, $a$ the orbital separation,  r$^2_{g}$ is the dimensionless gyration radius of the star\footnote{ r$^2_{g} = \frac{I}{MR^2}$, where $I$ the moment of inertia of the star, $M$ mass of the star,  $R$ radius of the star.}, and $\frac{K}{T}$, a coupling parameter depending on the tidal interaction mechanism, which we describe in the following section.

\subsection{Equilibrium tides}

For stars with an outer convective envelope, the turbulent viscosity on the equilibrium tides in the convective regions of a star is responsible for the dissipation of the tidal kinetic energy. In equilibrium tides, it is assumed that the star keeps the state of hydrostatic equilibrium, and all other dissipation processes are neglected. $(\frac{K}{T})_c$ is expressed \citep[see][and references therein]{2002MNRAS.329..897H} as:
    \begin{equation}\label{tsync1}
    \centering
    \left(\frac{K}{T}\right)_c = \frac{2}{21}\frac{f_{\rm conv}}{\tau_{\rm conv}}\frac{M_{\rm env}}{M} {\rm yr}^{-1},
    \end{equation}
where $f_{\rm conv}$ is a numerical factor and $\tau_{\rm conv}$ (in unit of year) the eddy turnover timescale \citep{1996ApJ...470.1187R}, and M$_{\rm env}$ the mass of the convective envelope.

\subsection{Dynamical tides}

For stars with outer radiative envelopes, radiative damping of the stellar oscillations is responsible for the dissipation of the tidal kinetic energy. This is also known as the regime of dynamical tides. In this regime $(\frac{K}{T})_r$ is defined as:
  \begin{equation}\label{tsync2}
    \centering
    \left(\frac{K}{T}\right)_r = \left(\frac{GM}{R^3 }\right)^{1/2}(1+q)^{5/6} E_{2} \left(\frac{R}{a}\right)^{5/2},
    \end{equation}
where E$_2$ (second order tidal coefficient, with higher orders being neglected) is a parameter that depends on the structure of the star and refers to the coupling between the tidal potential and gravity mode oscillations.

One widely used analytic approximation formula produced by \citet{2002MNRAS.329..897H}, based on tabulated results from \citet{1975A&A....41..329Z}, expresses E$_2$ as a function of the stellar mass:
	\begin{equation}\label{e21}
	\centering
	E_{2} = 1.592 \times 10^{-9} \left(\frac{M}{ M_\odot}\right)^{2.84}.
	\end{equation}
More recently \citet{2010ApJ...725..940Y} obtained the following expression:
	\begin{equation}\label{e22}
	\centering
	E_{2} = 10^{-1.37}\left(\frac{R_{\rm conv}}{R}\right)^8,
	\end{equation}
by fitting the dependence of E$_2$ on R$_{\rm conv}$/R, using the values given in Table 1 of \citet{1977A&A....57..383Z} for ZAMS stars with various masses.

There, R$_{\rm conv}$ denotes the radius of the convective core and $R$ is the radius of the star. The latter expression relates E$_2$ to the radius of the convective star and thus is more sensitive to the structure of the star. This formulation has been successfully implemented in several recent, detailed studies of rotation in massive stars \citep{2009A&A...497..243D,2013A&A...556A.100S,2018A&A...609A...3S}.

The original methodology to calculate E$_2$ was introduced by \citet{1975A&A....41..329Z}, and was discussed in more detail in later works \citep{1997A&A...318..187C,2013A&A...550A.100S,2017MNRAS.467.2146K}. Since both fitting formulae in Eqs. \ref{e21} and \ref{e22} for E$_2$ were calculated based on ZAMS, H-rich stellar models at solar metallicity, it is not obvious that they accurately represent He-rich stars over a variety of metallicities. We therefore decided to systematically investigate the dependence of  E$_2$, for both H-rich and He-rich stars, over a range of metallicities (i.e. 0.01 Z$_\odot$, 0.1 Z$_\odot$ and Z$_\odot$) and evolutionary stages.

For all the simulations of He-rich stars, we first create a naked He star at different masses. After that, with the same settings (i.e. stellar winds, rotational mixing parameters, Schwarzschild criterion for convection and overshooting with $\alpha_{ov}= 0.1$) as with H-rich stars in \S2, we compute the evolution of He-rich stars at different metallicities up to the central He exhaustion. The physical ingredients of the models used to compute E$_2$ are also the same as those used to compute the evolution of the He-rich stars in binary systems\footnote{The detailed list of parameters used for creation and the evolution of single He-rich stelar models can be found at http://mesastar.org/results.}. Appendix A provides the details of our method for calculating E$_2$ as well as a brief discussion.

In all cases, a functional form similar to the one adopted by Yoon et al.\ provides an adequate analytic approximation:
    \begin{equation}\label{newe2}
    {\rm E}_{2} = \begin{cases}
    10^{-0.42}\left(\frac{R_{\rm conv}}{R}\right)^{7.5} & {\rm H-rich\ stars} \\
    \\
    10^{-0.93}\left(\frac{R_{\rm conv}}{R}\right)^{6.7} & {\rm He-rich\ stars}
    \end{cases}
    \end{equation}
This updated relation is used in the expression for dynamical tides in the present work.

We note here that for fast rotating stars, it has been suggested that a high level of turbulence produced by rotation dominates over the radiative viscosity and hence equilibrium tides should be used despite the lack of an outer convective zone \citep{2007A&A...461.1057T,2008A&A...484..831D}. In the following section we test for the relative efficiency of equilibrium tides and dynamical tides in He-rich stars with radiative envelopes. However, we adopt the standard dynamical tides in all the detailed models that are presented on Section 5 and onward.

\section{Testing the efficiency of the tides in WR-BH binary systems}

Due to the large dimensionality of the available initial parameter space of WR-BH and WR-neutron star (NS) binaries (initial masses of the two binary components, initial orbital period, initial rotation of the He-rich star, and metallicity), it is computationally implausible to cover sufficiently densely the whole available parameter space. Knowing that tides play an important role only in close binary systems, we first do an order of magnitude test to identify the part of the parameter space where tides become relevant. We use MESA to evolve single He-rich stars with different metallicities, in the mass range 4 - 50 M$_\odot$ and steps of 2 M$_\odot$. The stellar structure information of these He-rich star models is used to calculate the tidal timescale of the synchronization with different compact object companions. For He-rich stars, the newly derived expression for E$_2$ from Eq. \ref{newe2} is adopted in the following calculations. Furthermore, we assume that a binary system has a He-rich star either with a BH of 10 or 30 M$_\odot$ or a NS of 1.4 M$_\odot$. Finally, we consider different initial orbital periods, $P$, spanning the range from 0.1 to 10 days. For each binary system, the Roche lobe radius, r$_L$, of the He-rich star provides a lower limit of the orbital period, as initially the He-rich star can not overfill its Roche lobe, where r$_L$ is given in units of the orbital separation by \citep{1983ApJ...268..368E}:
    \begin{equation}\label{eq9}
    \centering
    r_{\rm L} = \frac{0.49 q^{-2/3}}{0.6 q ^{-2/3} + ln(1+q ^{-1/3})} a,\quad  0 < q < \infty,
    \end{equation}
where $q$ is the mass ratio of the companion compact object to the He-rich star and $a$ the orbital separation. He-rich stars spend most of their lifetimes burning He in the core. We adopt the properties of the star half-way through its central He burning phase to calculate the synchronization timescale, T$_{\rm sync}$. The ratio of the synchronization timescale to the lifetime of the core He burning phase, T$_{\rm He}$, gives us a good handle on whether tides play a significant role in this binary configuration or not. In this approximation, we assume that the orbital separation remains constant during the whole evolution. In other words, we neglect the effects of stellar winds and spin-orbit angular momentum exchange.

\begin{figure*}[h]
\centering
\includegraphics[width=0.99\textwidth]{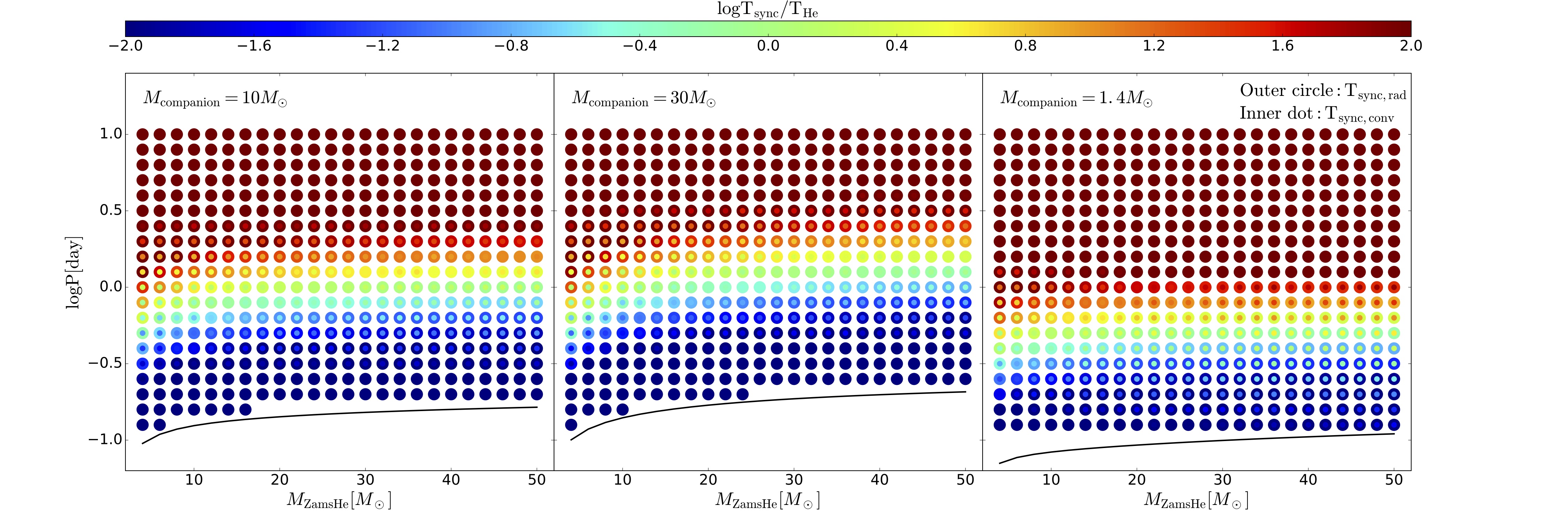}
\caption{The ratio of T$_{\rm sync}$/T$_{\rm He}$ as a function of the He-rich star initial mass and orbital period. Left panel: 10 M$_\odot$ BH as the companion, middle panel: 30 M$_\odot$ BH as the companion, right panel: 1.4 M$_\odot$ NS as the companion. Outer circle: T$_{\rm sync}$/T$_{\rm He}$ for dynamical tides, inner dot: T$_{\rm sync}$/T$_{\rm He}$ for equilibrium tides. The selected stellar structure profile for the calculation of T$_{\rm sync}$ refers to the stage when the central mass fractional He abundance is 0.5. The black line denotes the lower limit in orbital period, below which a He-rich star of a given mass would overfill its Roche lobe at ZAHeMS (Zero Age He Main Sequence: the time when a He star starts to burn He in the core, which is analog to ZAMS for core H burning). The metallicity of the He-rich star models shown in this figure is 0.01 Z$_\odot$, but the dependence of T$_{\rm sync}$/T$_{\rm He}$ on metallicity is very weak. Hence, two similar figures corresponding to the He-rich stars at 0.1 Z$_\odot$ and Z$_\odot$ are not shown here.}
\label{fig2}
\end{figure*}

\begin{figure*}[h]
\centering
\includegraphics[width=0.99\textwidth]{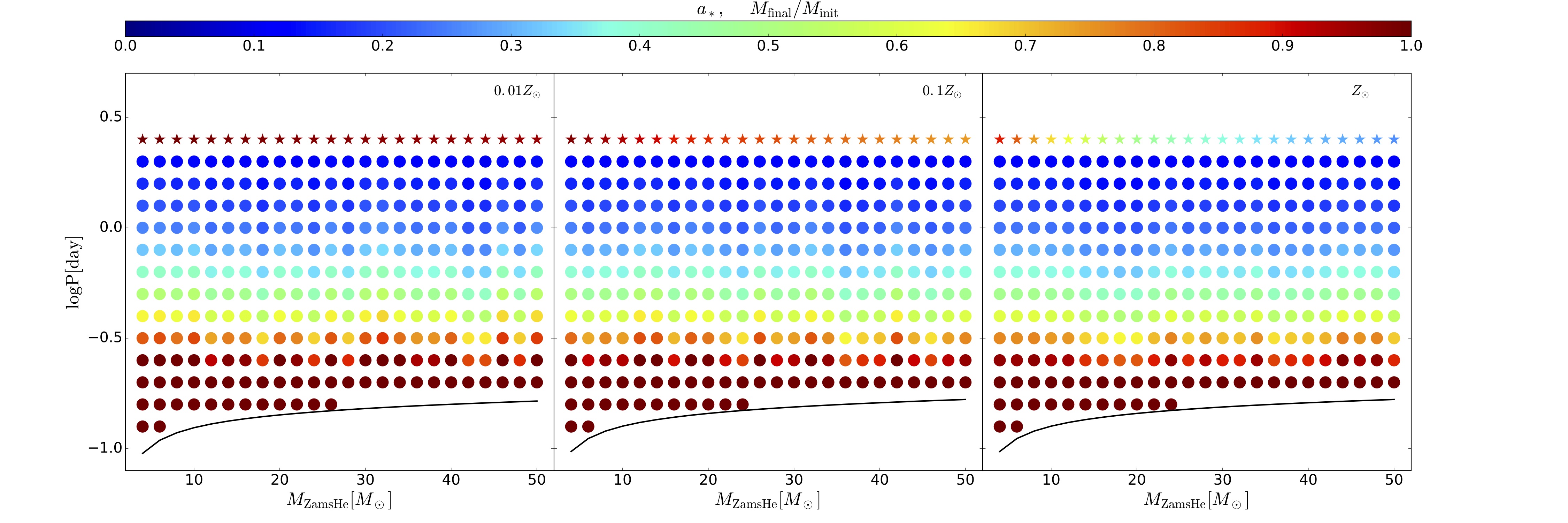}
\caption{Approximate estimate of the dimensionless spin $a_*$ of the resultant BH as a function of the He-rich initial mass and orbital period, denoted by the color of the filled circles. The star symbols, sharing the same color bar with $a_*$, refer to the ratio of the He-rich star's mass at the central carbon exhaustion to its initial mass, i.e. M$_{\rm final}$/ M$_{\rm init}$. For a given initial mass of the He-rich star, this quantity remains the same whatever the orbital period because in this estimation, stars are evolving as if they were single stars. Left panel: 0.01 Z$_\odot$, middle panel: 0.1 Z$_\odot$, right panel: Z$_\odot$. In these approximate estimates, the mass of the BH companion is assumed to be 10 M$_\odot$, while the orbital separation is assumed to remain constant. We also assume that tides instantaneously synchronize the spin of the He-rich star with the orbital angular velocity. The solid black line denotes the lower limit in orbital period, below which a He-rich star at a given specific mass would overfill its Roche lobe at ZAHeMS.}
\label{fig3}
\end{figure*}

The ratio T$_{\rm sync}$/T$_{\rm He}$ as a function of initial orbital period and the initial He-rich star mass, for three different companion masses (1.4, 10 and 30  M$_\odot$) is shown in Fig.~\ref{fig2}. The color of the outer circle and inner dot of each symbol corresponds to the T$_{\rm sync}$ estimated based on dynamical tides and equilibrium tides, respectively. Blue dots and circles correspond to systems in which tidal forces are expected to be relevant, while red dots and circles correspond to systems in which tides are likely to play a minor role in the binary's evolution.

Most importantly, we find that the strong dependence of the synchronization timescale on the ratio of stellar radius to orbital separation (T$_{\rm sync} \propto (R/a)^{-6}$) is the dominant factor, while the binary mass ratio and the exact dissipation mechanism is of less importance. Examining Fig.~\ref{fig2}, it is safe to say that for WR-BH and WR-NS binaries with orbital periods above $\sim$ 2 days, tides are not relevant, while binaries with periods below $\sim$ 0.3 days are expected to have He-rich stars with spins synchronized with the orbit. We note that in this approximation we find that metallicity has a negligible effect on T$_{\rm sync}$. The results for He-rich stars at higher metallicities (0.1 Z$_\odot$ and Z$_\odot$) are almost identical and thus not shown in the paper. Based on these estimates, we decided to limit the parameter space of initial orbital periods for which we will perform detailed calculations (see following sections) to below 2 days.

Stellar winds, scaled with metallicity, can greatly influence the final mass of the star, and this is clearly shown in Fig.~\ref{fig3}. The three panels correspond to He-rich stars at 0.01 Z$_\odot$, 0.1 Z$_\odot$ and Z$_\odot$, respectively. The color of the filled circles denotes the spin of the resultant BH. In this figure, the dimensionless spin $a_*$ is calculated based on the assumption that the He-rich star at the central He exhaustion still has enough mass to directly collapse to form a BH, and that the He-rich star is a solid body fully synchronized with the orbit. The color of the star symbols at the top of each panel refers to the ratio of the final to the initial He-rich mass. He-rich stars lose $\gtrsim3$ times more mass at solar metallicity compared with 0.01 Z$_\odot$.

From this figure, we see that for orbital periods below 2 days, the whole range of dimensionless spins (from 0 to 1) is covered. Fast spinning BH's are obtained only for short period systems, typically below 0.3 days. For orbital periods above about 1 day, BH spins are small, and for intermediate orbital period moderately spinning BH's are produced.

These numerical experiments however suffer from strong limitations. Principally, we assume that the orbital separation remains constant and that synchronization is instantaneous. In the next section, we compute more sophisticated models where the
effects of tides and of stellar winds on stellar rotation and orbital evolution
are consistently accounted for. Through tidal coupling, changes in the orbit and the stellar rotation then impact the He-star's evolution.

\section{Rotation of the second-born black hole}

Now that we have gained a qualitative understanding of which physical processes are important for binaries with different initial conditions and we have significantly limited the relevant part of the parameter space, we can go ahead and calculate grids of detailed calculations of close binaries consisting of a WR star and a compact object. The evolution of the binary is computed using the MESA code. The computation accounts for tidal coupling between the orbit and the He-rich star. Since the He-rich star has a radiative envelope, only the dynamical tide is considered. We assume that there is no MT between the He-rich star and the BH, which is assumed to be a point mass. More details about the computations can be found on the MESA web page \footnote{Detailed setting can be found at http://mesastar.org/results.}.

The initial conditions explored are the masses M$_1$ and M$_2$ of the two binary components, the initial rotation and metallicity of the WR star and the initial orbital period. For the He-rich stars, we cover the mass range from 4 to 48 M$_\odot$ with steps of 4 M$_\odot$ and the mass of the companion is assumed to be a NS of 1.4 M$_\odot$ or a BH of 10 or 30 M$_\odot$. The initial orbital periods are between 0.2 and 2 days. Below 0.2 days, the He-rich star overfills its Roche lobe at the onset of He burning, while for initial orbital periods above 2 days we showed in the previous section that tides are not important. Three metallicities (0.01 Z$_\odot$, 0.1 Z$_\odot$ and Z$_\odot$) are considered. Finally, the following initial rotations for the He-rich stars have been chosen: zero rotation,  angular velocity equal to the orbital angular velocity and 90\% of the critical angular velocity at the surface.

\begin{figure*}[h]
    \centering
    \includegraphics[width=0.99\textwidth]{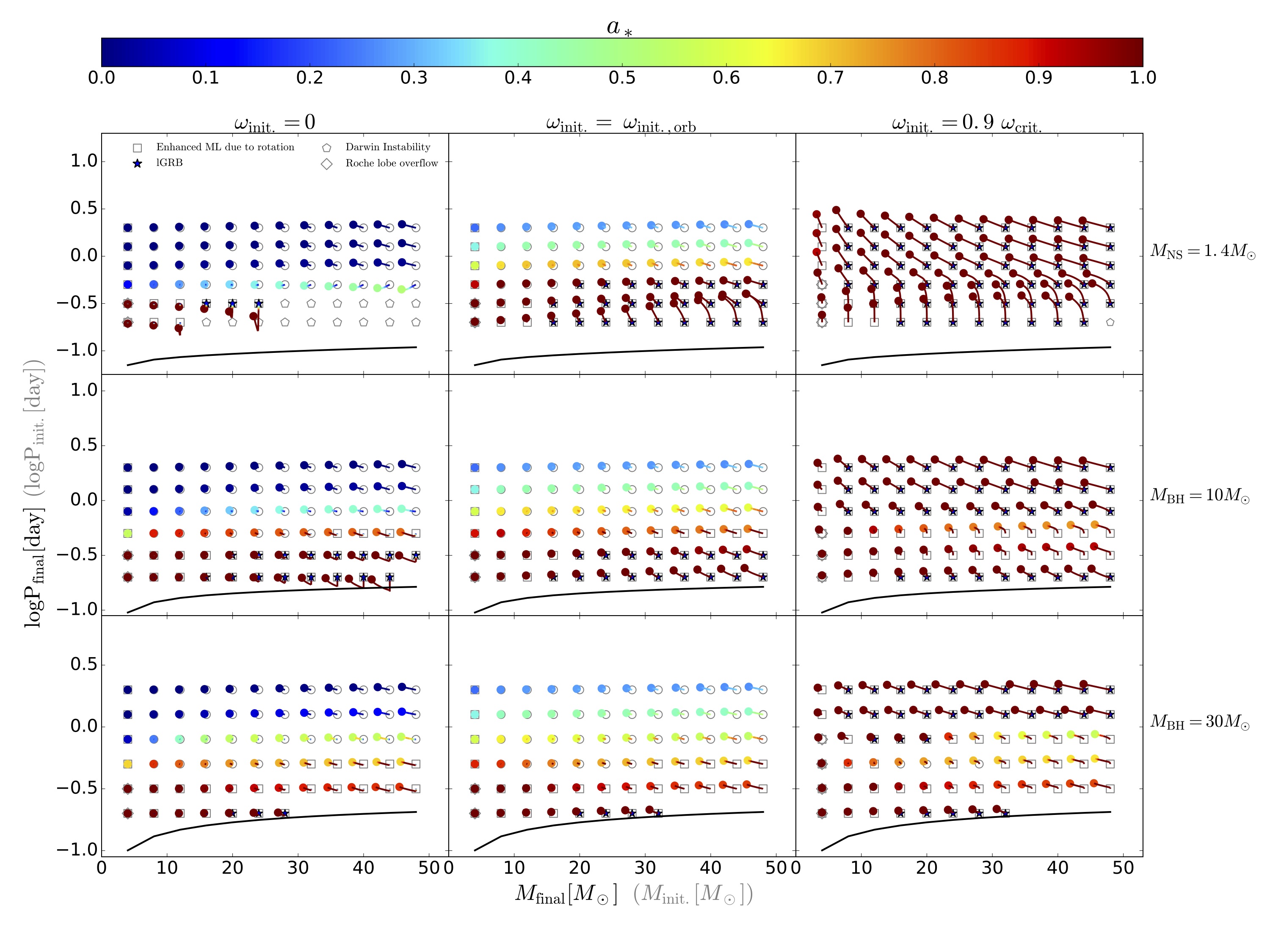}
    \caption{Spin parameter $a_*$ (see the color bar at the top) as a function of the orbital period and masses of the He-rich stars. The metallicity of all  He-rich stars is 0.01 Z$_\odot$. The grey symbols indicate the initial conditions and the color symbols indicate the final ones. The colored lines linking these two symbols show the evolution of the binary. The color along the line gives $a_*$ along the evolution. Black solid lines refer to the lower limit of the orbital period. At that limit, the He-rich star fills its Roche lobe at the beginning of the core He burning phase. Square: models for which rotation increases the mass lost by 10\% with respect to the mass lost by non-rotating models; pentagon: Darwin instability; star: lGRB; Diamond: He-rich star starts to fill its Roche lobe. The three columns correspond to different initial velocities of the He-rich stars and the three rows correspond to different masses for the companion.  All the He-rich stars have a metallicity equal to 0.01 Z$_\odot$.}
    \label{fig4}
    \end{figure*}

In Fig.~\ref{fig4}, we show the spin $a_*$ of the second-born BH as a function of the He-rich star's mass and the orbital period for a metallicity Z = 0.01 Z$_\odot$. Figures 5 and 6 present the corresponding trends for metallicities 0.1 and 1.0 Z$_\odot$, respectively. In each figure, the three columns correspond to the different initial angular velocities of the He-rich star, i.e., $\omega_{\rm init} = 0$, $\omega_{\rm init} = \omega_{\rm init,orb}$ and $\omega_{\rm init}$ = 0.9 $\omega_{\rm crit}$. The three rows correspond to the different masses of the compact object companion, i.e., 1.4, 10 and 30 M$_\odot$. The color bar denotes the spin of the second-born BH under the assumption that He-rich stars at the end of their evolution directly collapse to form BHs, conserving angular momentum.  The black solid lines in each panel indicate the initial orbital period as a function of M$_{\rm ZAHeMS}$ below which the He-rich star overfills its Roche lobe on the ZAHeMS. Different empty grey symbols indicate various evolutionary features, i.e., grey squares: enhanced mass loss due to rotation (i.e. at least an increase by 10\% of the total mass lost compared to the total mass lost by the corresponding non-rotating model); pentagon: Darwin instability \citep{1879RSPS...29..168D}, this instability occurs when the star evolves into a stage where its spin angular momentum exceeds 1/3 of the orbital angular momentum. In that case the binary system rapidly merges; star: lGRB can be formed during the direct core collapse; diamond: He-rich star evolves to fill its Roche lobe.

     \begin{figure*}[h]
    \centering
    \includegraphics[width=0.99\textwidth]{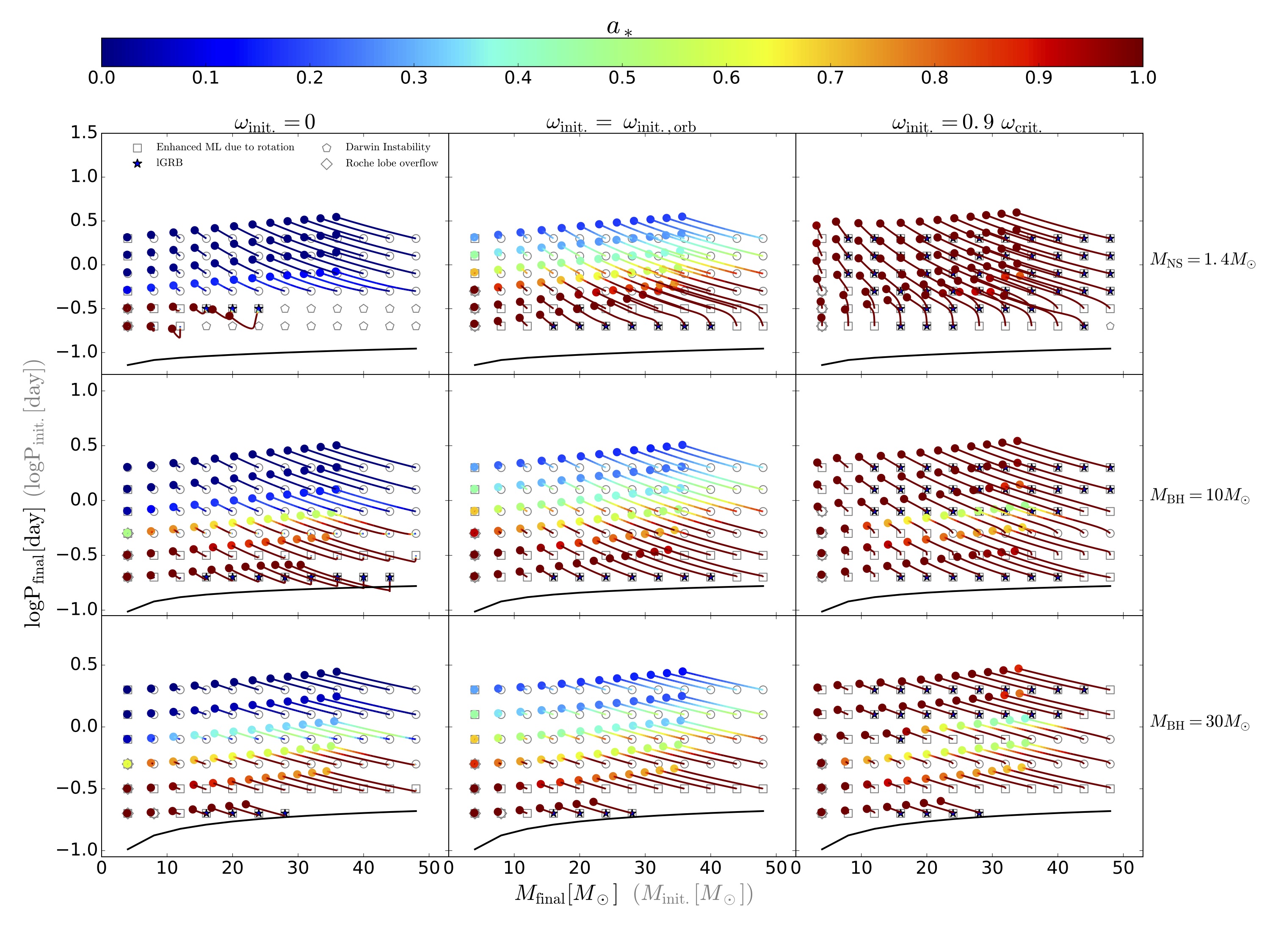}
    \caption{The same as Fig.~\ref{fig4}, but for the metallicity Z= 0.1 Z$_\odot$.}
    \label{fig5}
    \end{figure*}

    \begin{figure*}[h]
    \centering
    \includegraphics[width=0.99\textwidth]{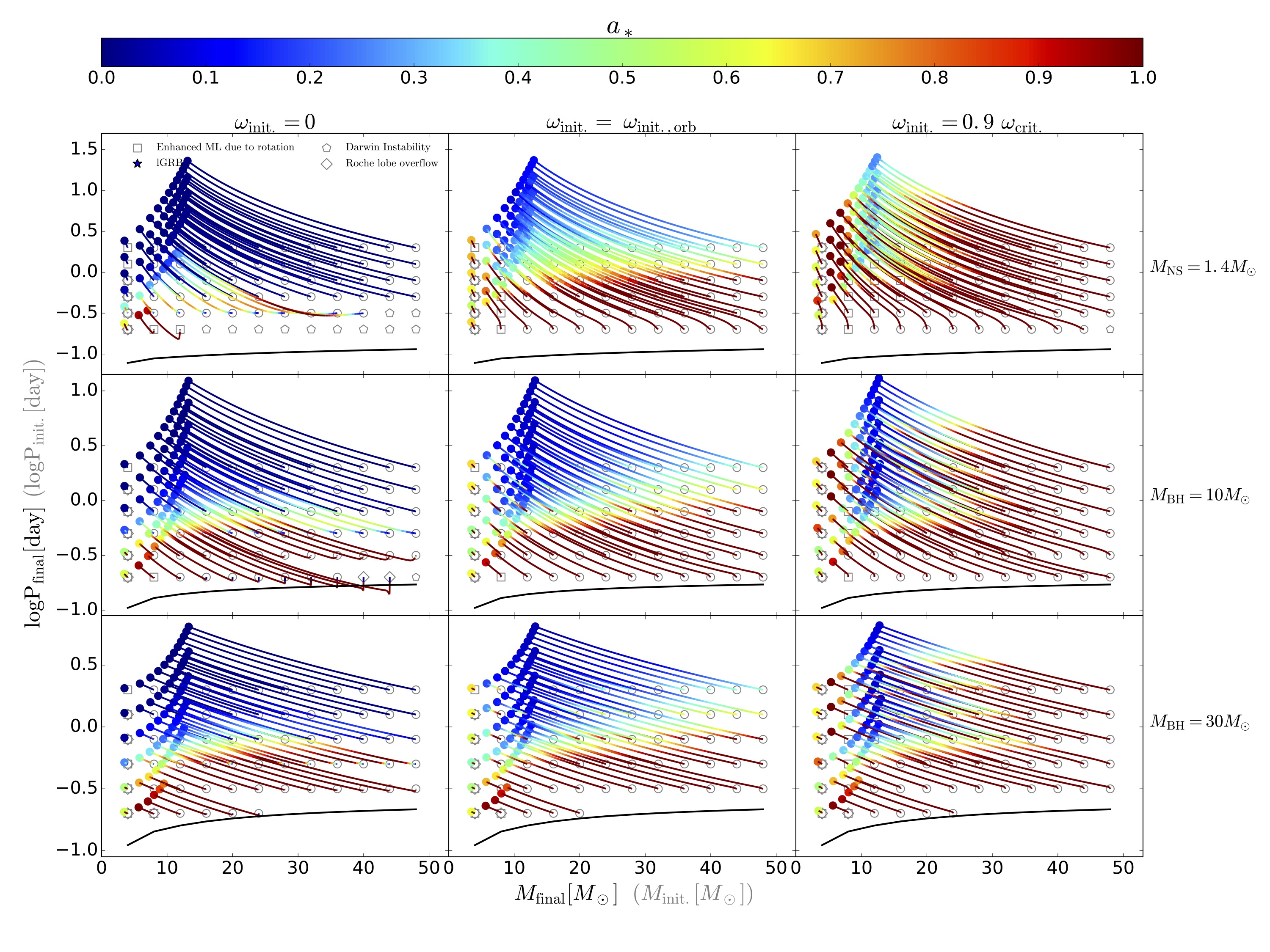}
    \caption{The as the Fig.~\ref{fig4}, but for the metallicity Z= Z$_\odot$.}
    \label{fig6}
    \end{figure*}

\subsection{Dependence of orbital evolution and He-star rotation on different binary properties}

Before describing the results, let us remind a few general trends: first tides tend to equalize the rotation period of the star and the orbital period. This effect implies that when the star has a relatively longer rotation period (or a slower rotation rate) compared with the orbital period, then tides tend to transfer angular momentum from the orbit to the star spinning it up. Therefore, the orbit shrinks. On the contrary, when the spin period of the star is shorter than the orbital one, angular momentum is transferred from the star to the orbit. Consequently, the star spins down, the orbit widens and the orbital period increases. Second, mass loss has counteracting effects. On the one hand, it decreases the mass of one component, thus its gravitational attraction, which widens the orbit.
On the other hand, it removes orbital angular momentum, shrinking the orbit. Under standard assumptions of ``fast'' stellar winds the overall effect is the expansion of the orbit.
Mass loss also removes spin angular momentum from the star tending to spin down the star. For systems near spin-orbit synchronization, this implies that tides will spin up the primary by transferring angular momentum from the orbit to the star. This also tends to shrink the orbit.

In addition to mass loss, structural changes of the star as a function of time modify the tidal interaction and thus contribute in modifying the orbit, too. As it can be guessed from this discussion, it is not easy, without performing detailed calculations, to know what will be the evolution of such systems. Depending on which effect dominates, the angular momentum of the He-rich star may increase or decrease.

\subsubsection{Dependence on initial orbital period}

Let us begin describing the upper left panel of Fig.~\ref{fig4}. Tides are weak at large orbital periods. Thus no spin-up occurs (starting with a low rotation implies that tides can only spin up the primary) and the final spin of the He-rich star and of the resulting BH remain low. For higher masses, mass loss slightly decreases the mass and widens the orbit, hence the evolution towards the upper left in the period-mass plane.

At an orbital period around half a day (log $P\approx -0.3$ ), tides become important and spin up the He-rich star. In this case however, the synchronization timescale is still comparable to or longer than the He-star lifetime and thus the binary never quite reaches a state of synchronization. Instead, the final rotation of the WR star is somewhere between its initial value and the one corresponding to the orbital angular velocity. The final spin of the BH is between 0 and 0.5.

At still smaller orbital periods, tides are efficient enough to make the He-star rapidly reach a rotation rate that is equal to the angular orbital velocity. At the same time, the angular momentum that is transferred from the orbit to the star in order to spin it up results in the initial shrinkage of the orbit. This phase corresponds to the nearly vertically downward evolution. This initial synchronization phase is short and hence mass losses have no time to significantly change the mass of the binary. The orbital period decreases because angular momentum is transferred from the orbit to the star. Once synchronization is reached, the rotation of the primary is maintained near the synchronized value by the tidal interaction. The orbit widens again because the mass loss
term dominates the tidal one in the evolution of the orbital distance.

\subsubsection{Dependence on initial rotation of the He star}

Let us now see how the results change when higher
initial rotations are considered (see the upper middle panel in Fig.~\ref{fig4} which shows the case when
the initial rotation of the He-rich star is synchronized with the orbit). We see that starting from a higher initial rotation rate for the He-rich stars produces at the end faster rotating BHs. BH spins are found in the range between 0.3 and 1.0. We also see that the orbital period always increases,
and thus the orbit becomes wider. For the large orbital periods, for which tides are weak, this is an effect of mass loss that decreases the mutual attraction between the two bodies and thus widens the orbit.
The star's spin is slightly slowed down, too, due to mass loss (one sees that the beginning of the line has a colour corresponding to a higher rotation than its end). More interestingly, and in contrast with the results for initially non-rotating He-rich stars, we see that even for small initial orbital periods the orbital period increases. Since here we start from synchronization, tides do not initially transfer a significant amount of angular momentum from the orbit to the stars, and hence from the beginning of the evolution the effect of mass loss dominates. Tides counteract the spin down of the star due to mass loss and allow the star to keep a fast rotation.

Increasing further the initial velocity up to 90\% the critical one (see the upper right panel), produces only fast rotating BHs for all the initial conditions explored in this plot. In that case, spins of the BHs are always near 1.0. The evolution always tends to increase the orbital period. This widening of the orbit results, as before, from mass loss. For shorter periods though, where tides are efficient, it may also come from the fact that the tides, before synchronization, slow down the star and thus transfer angular momentum from the star to the orbit causing it to widen.

\subsubsection{Dependence on the mass of the compact-object binary companion}

Let us now see how the results change when the mass of the compact object is varied. If we consider systems where the orbital periods are kept fixed, increasing the mass of the compact object increases the orbital separation $a$ ($a$ varies as $(1+q)^{1/3}$). On the other hand the quantity $1/T_{\rm sync}$ varies as $q^2/(1+q)^2$ and thus increases when the mass ratio increases. This indicates that the tides at a given orbital period are stronger (despite the increase of the distance) when the mass of the compact object is larger.

For the case of low initial rotation (compare the left middle panel to the left upper one in Fig.~\ref{fig4}), and considering a fixed initial orbital period and mass for the He-rich star, increasing the mass of the compact object more efficiently spins up the He-rich star and thus produces fast spinning black holes starting from longer initial orbital periods.

Comparing now the bottom left panel with the middle left one, i.e. passing from a 10 to a 30 M$_\odot$ BH, increases the spin of the second-born BH for longer periods, but slightly decreases the spin for the shorter ones. This appears a bit counter-intuitive at first, since one would expect that increasing the BH mass would always more efficiently spin up the He-rich star. However, as we increase the mass of the compact object companion, from 1.4M$_\odot$ all the way to 30M$_\odot$, the ratio of the spin angular momentum of the He-rich star to the orbital angular momentum, in a state of synchronization, decreases. This means that a smaller fraction of the orbital angular momentum has to be transferred to the He-rich star to spin it up from an initially low rotation to synchronization, and hence the orbit will shrink less in the initial phase until the system is brought into synchronization. To highlight this effect, let us consider the case of a 48 M$_\odot$ He-rich star with a 10 M$_\odot$ BH and an initial orbital period of 0.3 days (log $P$=-0.5). Initially, due to the evolution towards synchronization, the orbit shrinks. This produces the small evolution downwards (a bit to the left because of the mass loss).  After the binary reaches synchronization,  the orbit widens again because the effects of mass loss. For more massive compact-object companions, this initial phase towards synchronization leads to a negligible shrinking of the orbit, as a much smaller fraction of the orbital angular momentum needs to be transferred to the He-rich star, and as a result the final orbital period of the binary, at the point of carbon exhaustion, is longer.

Despite the slightly shorter final orbital periods for higher mass compact-object companions, if we compare the final spins of the resulting second-born BHs, we see that the final spins decrease with increasing companion mass, going from a value near 1 for the smallest mass companion to a value near 0.8 for the most massive one. What is the explanation for this trend? This behavior is due to the fact that at the very end of the core He-burning phase, the entire star quickly contracts. This contraction de-synchronizes the star from the orbit since the contraction timescale is shorter than the tidal timescale. After the contraction, the star is spinning faster than the orbit, with tides acting to slow down the star. The more massive the companion is, the stronger the tidal coupling, and thus also the more efficient the spin down. Note that this fast contraction also occurs for He-rich stars in systems with larger orbital periods. However, in these systems the loss of spin angular momentum due to stellar winds throughout the evolution of the He-star is not compensated by tides, which are too weak. When the star contracts at the end of the core He-burning phase, it has too little angular momentum to reach large spins.

When one starts from a configuration where the binary is synchronized, increasing the mass of the compact-object companion has two effects at short orbital periods (compare the panels in the middle column in Fig.~\ref{fig4}). Firstly, it decreases the widening of the orbit, and secondly it tends to produce slower rotating BHs in the regions where tides are important. The latter effect is explained above, while the former results from the two counteracting effects of the mass loss. On the one hand, mass loss spins down the He-rich star and forces tides to continuously transfer angular momentum from the orbit to the star. On the other hand it reduces the mass of the He-rich star and  tends to widen the orbit. The net effect, i.e. the widening of the orbit, remains the same when the mass of the compact object increases. However, since in the case of a higher mass compact-object companion, a smaller fraction of the total binary mass is lost in winds, the overall expansion of the orbit is smaller. As a reminder, under the assumption that the mass lost is carrying the specific angular momentum of the mass-losing star (Jeans mass loss), the ratio of the final to the initial orbital separation is inversely proportional to the ratio of the final to the initial total binary mass (a$_{\rm final}$/a$_{\rm initial}$=M$_{\rm binary,initial}$/M$_{\rm binary,final}$).

Starting with still higher initial rotations for the He-rich stars (see the right column in Fig. 4)  produces in general faster rotating BHs. We note the same behaviour as for the cases shown in the middle column, namely that increasing the mass of the compact remnant produces smaller BHs rotations in some initial mass and period ranges.

\subsubsection{Dependence on the metallicity of the He star}

When the metallicity increases (see Fig. 5 and 6), the same qualitative behaviours are obtained but the effect of mass loss dominates the evolution in almost all cases. In the period-mass diagrams, stronger mass losses bring the star to smaller final masses and longer orbital periods. At solar metallicity, even when starting with a high initial rotation for the He-rich star, most of the cases studied here end with slowly rotating BHs. The only exception is for the least massive He-rich stars considered here, for which the mass loss is much less important.

\subsubsection{Summarizing the effects of different initial properties on the evolution of the binary}

These computations show how the effects of mass loss and tides impact the final spin of the second-born compact object. The following results have been obtained:
\begin{itemize}
\item Independent of the initial rotation of the He-rich star and its
metallicity, fast spin at the end of the evolution ($a_* > 0.9$) is obtained for short orbital systems, below about 0.3 days, and for initial masses below about 30 M$_\odot$. In those systems, tides are the key players in determining the final spin.
\item For orbital periods above about 0.3 days and
at low metallicities, the initial rotation of the He-rich star is the main factor
impacting the final spin. The faster the initial rotation is, the faster the final BH spin.
\item For orbital periods above about 0.3 days and
at solar metallicity, stellar winds have a major impact on the final rotation for stars with masses above about 25 M$_\odot$.
Mass losses in these cases efficiently slow down the He-rich star and modest final spins are obtained.
\item The mass of the compact-object companion has an impact only in cases in which tides are sufficiently strong, i.e. for orbital periods below about 1 day. In general a more massive companion produces a smaller final spin. This comes from the fact that when the star contracts at the end of the core He-burning phase, and thus spins faster and faster, tides tend to slow it down. The more massive the compact-object companion is, the more efficiently the He star slows down.
\end{itemize}
As was already envisioned from the order of magnitude estimates presented in Sect.~4, the whole range of final spins can be reached for a given He-rich star at low metallicity
depending on the initial orbital period and rotation (the mass of the compact object has little influence on the range of values that can be reached). At solar metallicity and
for the most massive stars, this statement is no longer true. For these stars, only
low spin parameters are obtained independently of the initial orbital period, the initial rotation or the mass of the companion.

\subsection{Mass loss enhanced by rotation}

Squares in Figs 4, 5 and 6 indicate  rotating models in which the total mass lost is more than 1.1 times the total mass lost by the corresponding model without rotation. At low initial rotation and for Z=0.01 Z$_\odot$, squares appear for small orbital periods, {\it i.e.} for those cases where tides are efficient enough to spin-up the star. For faster initial rotation, squares, obviously, cover a larger zone of the period-mass diagram.

Comparing models with binary component masses, metallicity and initial rotation, but with different initial orbital periods, where in some cases the stars experience enhanced mass loss while in others they don't, we see that the final He-rich star masses are not significantly different. Hence, we infer that these enhancements in mass loss should also have little effects on the final rotation as well as the orbital evolution.

\subsection{Systems with Mass Transfer}

Diamonds indicate those systems encountering the Roche limit during their evolution. Only models with initial M$_{\rm ZAHeMS}$ = 4 M$_{\odot}$ and initial orbital periods of 0.5 days or less, overfill their Roche lobes. This is because low-mass He-rich stars ($\lesssim$ 4 M$_\odot$) expand towards the end of their evolution.

However, since this mass transfer occurs at the very end of the evolution, the effects on the evolution of the binary and the final angular momentum of the second-born BH progenitor are negligible. We performed tests where we compared results obtained with and without accounting for this mass transfer and the differences concerning the spin of the second-born BH are very small and thus this effect is negligible.

\subsection{The Darwin instability}

Pentagons indicate that a Darwin instability is encountered. This occurs only for the tightest systems and for those systems with an initial low rotation except for one case (systems with a 1.4 M$_\odot$ NS with an initial rotation 0.9$\omega_{\rm crit}$ for all the metallicities considered here). This instability requires that a large amount of the orbital angular momentum be transferred into the spin angular momentum of the He-rich star. Obviously this can only occur for tight systems because only tides can transfer angular momentum between the orbital and the spin angular momentum reservoirs. The transfer from the orbital to the spin reservoir is the most efficient when the difference between the low spin and the high orbital spin is the greatest, which means for the cases that start from a low initial rotation rate.
The conditions are more favorable for low mass compact stars because the orbital angular momentums in these systems are the lowest.

The domain where the Darwin instability is reached disappears at high initial rotations. This is because when one starts from a high rotation, tides tend to slow down the star and thus to transfer angular momentum from the star to the orbit making the system evolve away from the conditions needed for this instability to occur. There is however, as already indicated above, one exception: systems with a 1.4 M$_\odot$ NS and an initial rotation of 0.9$\omega_{\rm crit}$. In this case, from the beginning of the evolution, the spin angular momentum of the star is larger than 1/3 the orbital angular momentum. This is therefore not an evolutionary effect but rather, it is due to the initial binary configuration.

\subsection{The long gamma ray bursts}

He-rich stars that are potential progenitors of lGRBs are shown by a star in Figs 4, 5 and 6. According to the collapsar model, lGRBs are formed on the condition that enough kinetic energy is available to launch a jet during the core collapse from massive stars \citep{1993AAS...182.5505W}.
In this work, we follow the procedure suggested in \citet{2006A&A...460..199Y} to decide whether the collapse of the core would produce a lGRB or not. More specifically, a lGRB is produced if any part of the carbon/oxygen core has a specific angular momentum larger than the one at the last stable orbit j$_{\rm LSO}$ around a black hole with a mass equal to the enclosed mass of the specific shell  \citep{1972ApJ...178..347B,1973blho.conf..343N,2007A&A...465L..29C,2008A&A...484..831D,2013ApJ...767L..36W}.

At Z=0.01 Z$_\odot$, and for low initial rotation periods (see upper left panel of Fig.~4), the domain of the lGRBs is quite limited to the most extreme cases, i.e., those suffering the strongest tidal interactions while not encountering the Darwin instability. Too low initial mass models produce a neutron star and thus are discarded as possible progenitors of lGRBs. The conditions favorable for lGRBs in the period-mass diagram are in general more extended in the case of faster initial rotations. This is expected since the reservoir of spin angular momentum is larger. Looking at the middle and right columns of Fig.~4, we also note that the domain for lGRB reduces when the mass of the compact object increases. Also, in the case, for instance, of a 30 M$_\odot$ BH, the most favorable cases are in the upper and lower parts of the orbital range considered here. When the metallicity increases, the extent of the initial parameter space leading to lGRBs generally reduces and nearly completely disappears at solar metallicity.

Overall, we see that the most favorable conditions for obtaining lGRBs from close binary systems are a high initial rotation for the He-rich star, a low mass compact-object companion and a low metallicity.

\section{Merging timescales}

\begin{figure*}[h]
    \centering
    \includegraphics[width=0.99\textwidth]{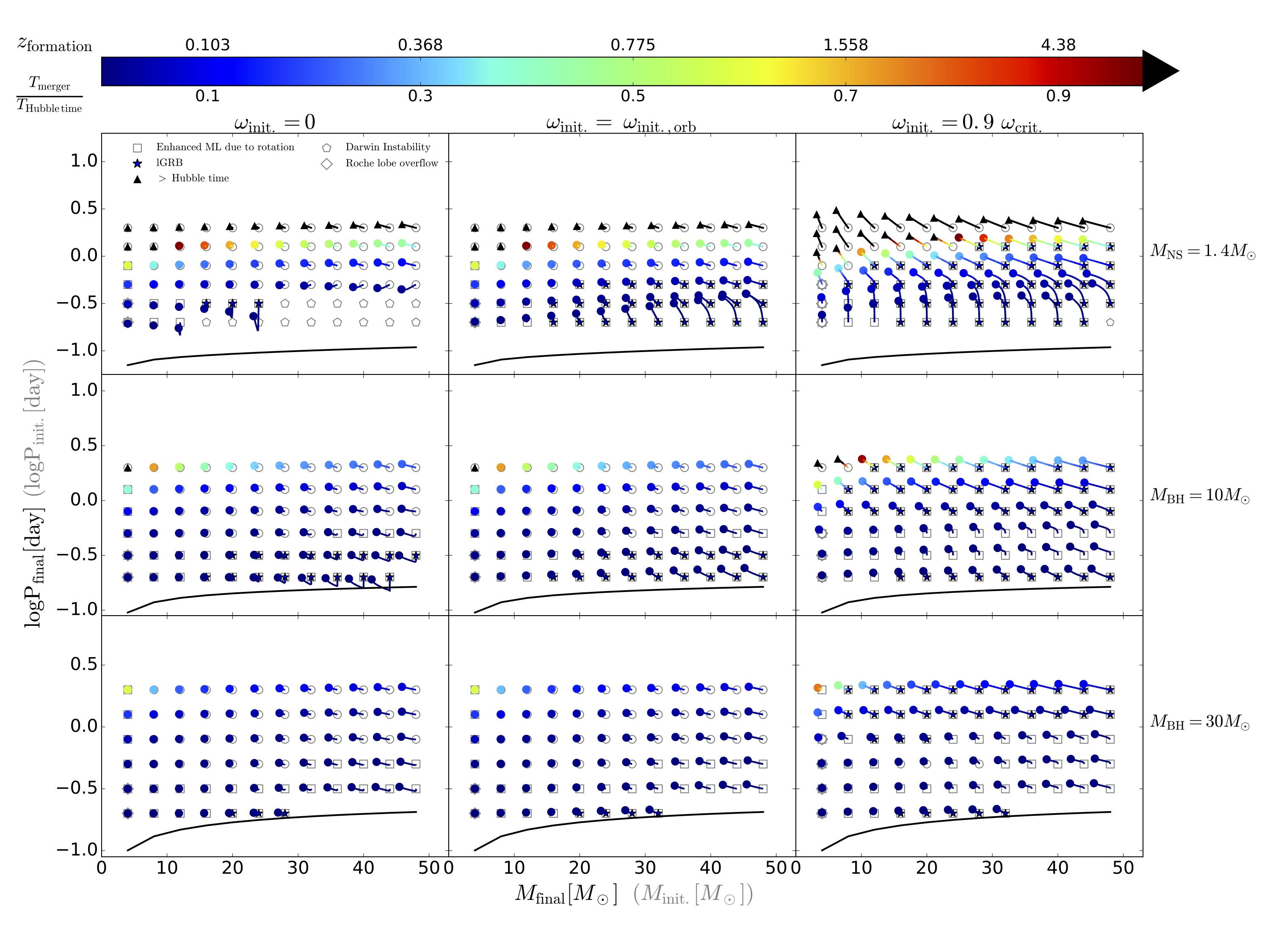}
    \caption{The same as Fig.~\ref{fig4}, but the color bar refers to the merger timescale of the two compact objects due to gravitational wave emissions. Black triangles refer to the systems whose merger timescales are longer than the present age of the Universe ($\sim 13.8$ $\rm Gyr$). z$_{\rm formation}$ refers to the redshift of the formation of the binary compact object, assuming that the merger took place at redshift$\sim$0 and adopting the standard cosmological parameters \citep{2016A&A...594A..13P}.}
    \label{fig7}
    \end{figure*}

After the second-born BH forms, gravitational wave (GW) emission removes angular momentum from the orbit of the two compact objects, shrinking it, and  eventually leading to the merger of the two compact objects. The timescale for  the merger of a binary compact object due to GWs is given by~\citep{1964PhRv..136.1224P}
    \begin{equation}\label{eq11}
    \centering
    T_{\rm merger} = \frac{5}{512}\frac{c^5}{G^3M^3}\frac{2 q^{-2}}{1+q^{-1}}a^4,
    \end{equation}
where $M$ is the mass of the second-born BH, $q$ the mass ratio of the companion to the second-born BH and $a$ is the orbital separation. In Fig.~\ref{fig7}, the color bar indicates T$_{\rm merger}$ due to GW emission assuming that the He-rich star at the end of its evolution can collapse directly to form a BH. Figures 8 and 9 are similar to Fig.~\ref{fig7}, but correspond to metallicities of 0.1 Z$_\odot$ and Z$_\odot$, respectively. When the merging timescale is for instance equal to 10\% the age of the Universe, and assuming that the merger of this binary is observed today in the local universe (i.e. $z_{\rm observed}\sim 0$), this binary compact object must have formed at redshift $z_{\rm formation}=0.103$\footnote{In this paper, we adopt the standard spatially flat $\Lambda$CDM cosmology with Hubble constant H$_0$ = 67.8 km s$^{-1}$ Mpc$^{-1}$, a matter density parameter $\omega_{\rm m} = 0.308$ and vacuum density parameter $\omega_{\Lambda} = 0.692$ \citep{2016A&A...594A..13P}, to calculate the redshift corresponding to the timescale of merger events.}. In all these three figures, black triangles refer to the systems whose merging timescale  is longer than the Hubble time ($\sim 13.8$ $Gyr$).

    \begin{figure*}[h]
    \centering
    \includegraphics[width=0.99\textwidth]{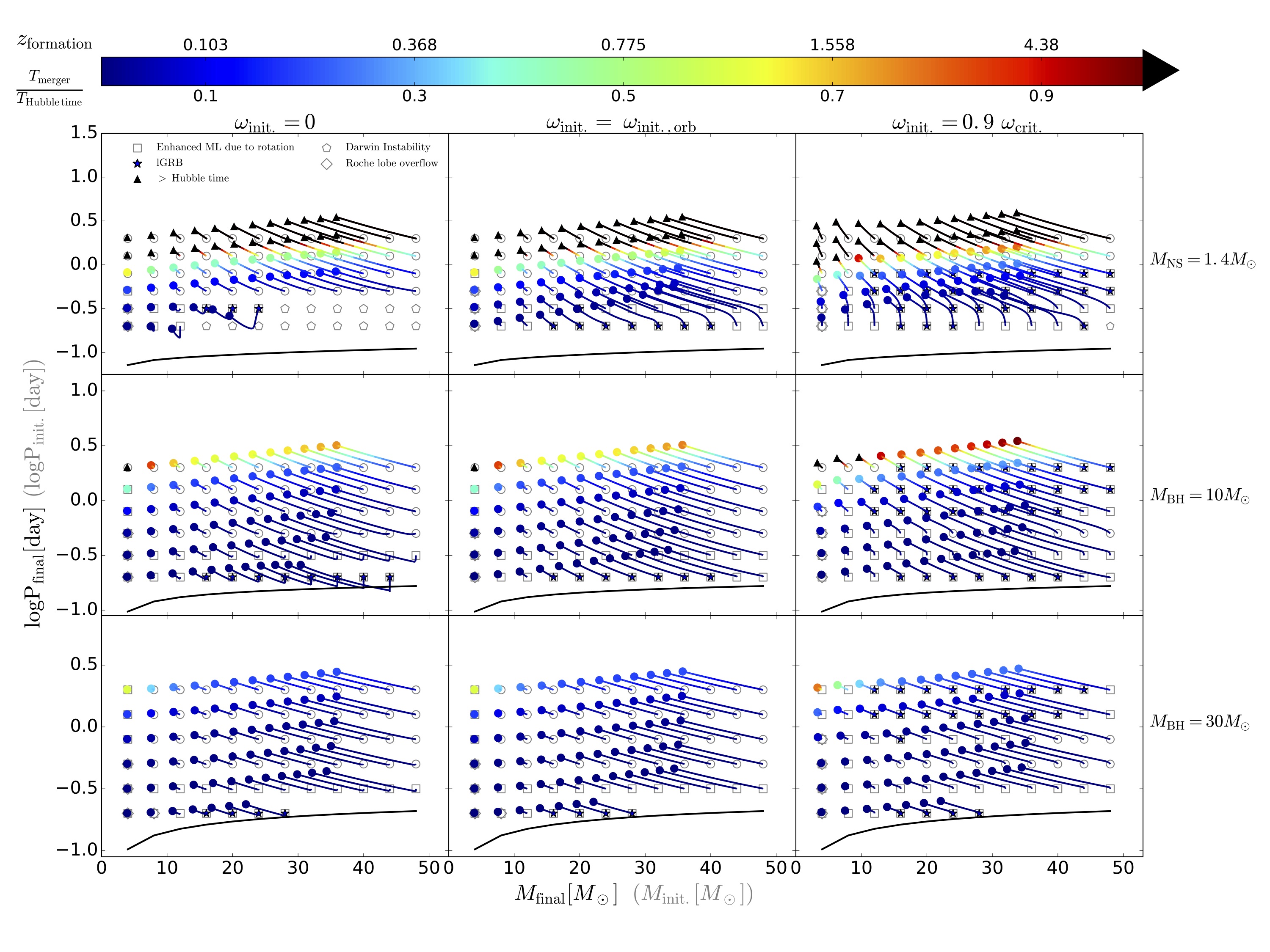}
    \caption{The same as Fig.~\ref{fig7}, but for the metallicity Z= 0.1 Z$_\odot$.}
    \label{fig8}
    \end{figure*}

    \begin{figure*}[h]
    \centering
    \includegraphics[width=0.99\textwidth]{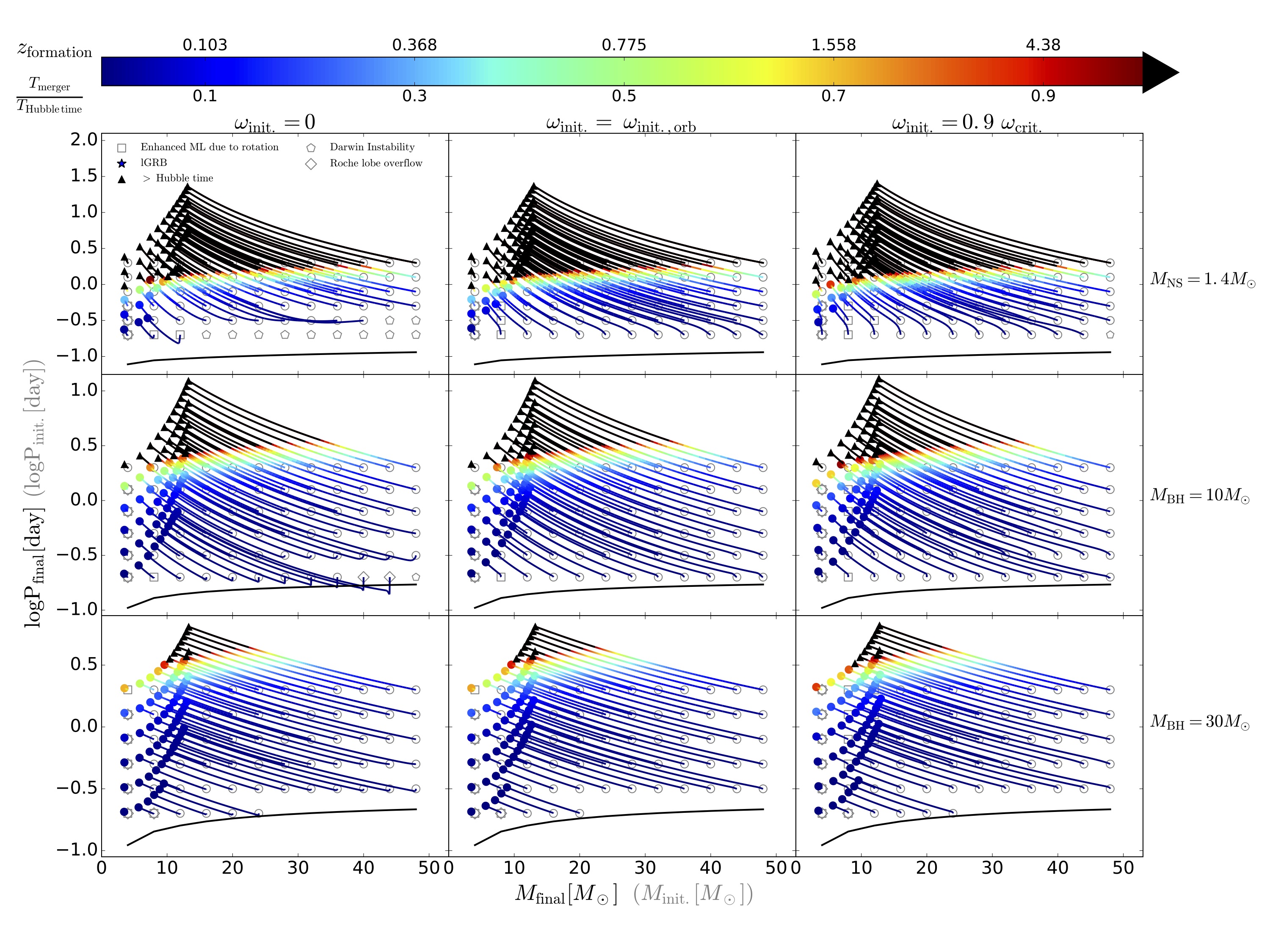}
    \caption{The same as Fig.~\ref{fig7}, but for the metallicity Z=  Z$_\odot$.}
    \label{fig9}
    \end{figure*}

The initial orbital separation $a$, or the initial orbital period, is the most important factor in determining T$_{\rm merger}$ (T$_{\rm merger} \propto a^4$). The mass of the He-rich star is also an important parameter. Decreasing the mass of the He-rich star, and keeping all other parameters equal, makes the merging timescale longer (T$_{\rm merger} \propto M^{-3}$). Due to this dependence on orbital period and the mass of the He-rich stars, merging timescales shorter than the Hubble time are obtained for small initial orbital periods and/or massive He-rich stars. At high metallicity (Fig.~9), one can see clearly that the upper limit of the initial orbital period below which merging timescales are inferior to the Hubble time increases when the mass of the compact companion increases. This also occurs for the lower metallicities but is less apparent in the figures. This is a rather obvious consequence of the fact that increasing the mass of the compact remnant implies stronger tides and thus shorter final orbital periods. One also notes that at high metallicity, decreasing the initial mass of the He-rich star, starting from a given initial orbital period, decreases the you merging timescale. This is likely due to the fact that lower-mass He-rich stars lose less mass by stellar winds. This in turns  implies less widening of the orbit and thus stronger tidal forces.

\begin{figure*}[h]
    \centering
    \includegraphics[width=0.99\textwidth]{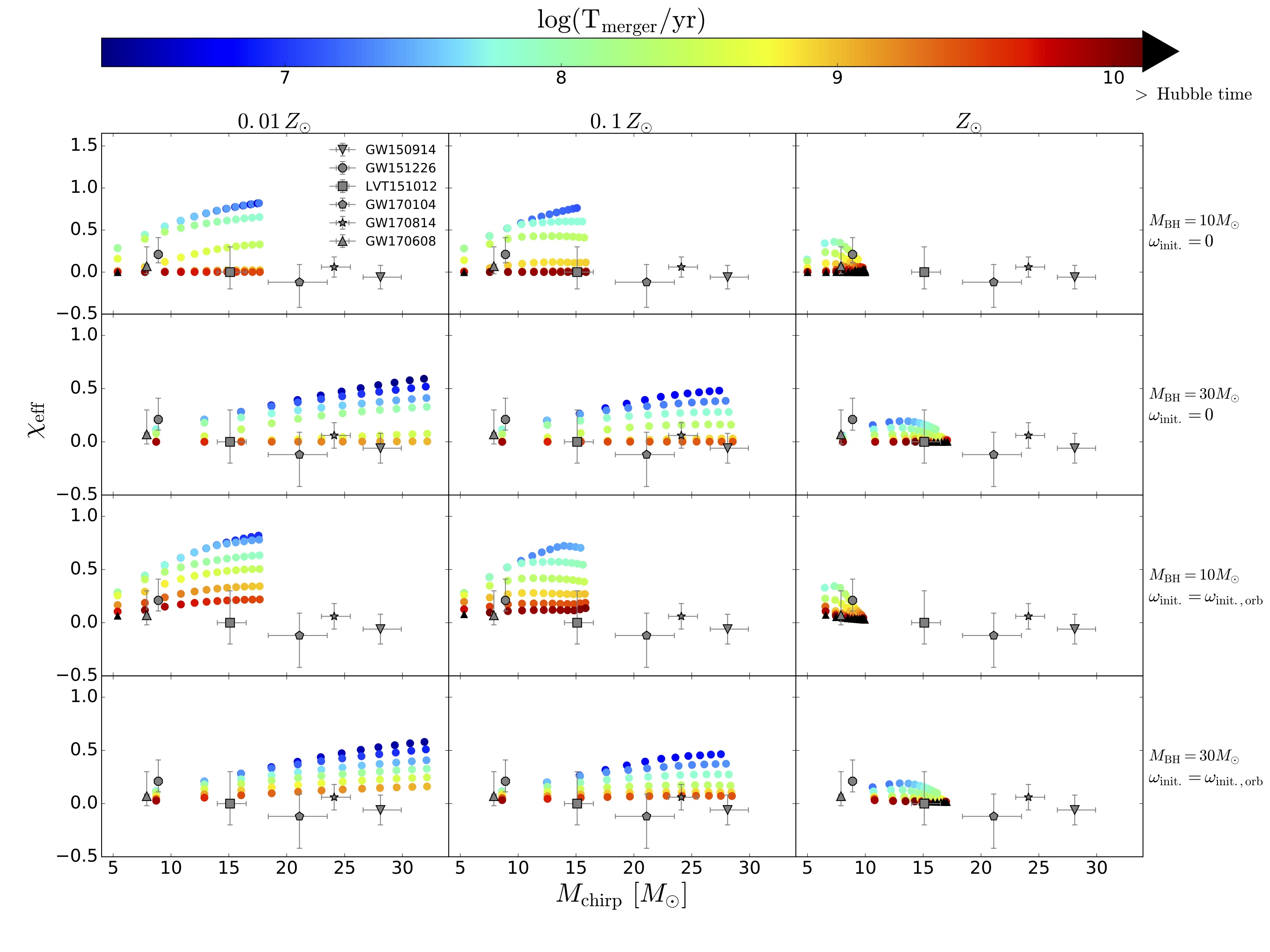}
    \caption{T$_{\rm merger}$ as a function of $\chi_{\rm eff}$ and M$_{\rm chirp}$. The first row including three panels corresponds to the binary systems in which the companions are 10 M$_\odot$ BHs and there is no initial rotation for the He-rich stars. The other three rows are similar to the first row, but with different companions and initial rotation of the He-rich stars. The three columns refer to different metallicities of the He-rich stars, i.e., 0.01 Z$_\odot$, 0.1 Z$_\odot$ and Z$_\odot$. The colored dots correspond to the T$_{\rm merger}$. Three gravitational events and one candidate with specific $\chi_{\rm eff}$ and M$_{\rm chirp}$ from the observation of AdLIGO are shown on each panel. Black triangles correspond to binary systems whose T$_{\rm merger}$ is longer than the Hubble time.}
    \label{fig10}
    \end{figure*}

Interestingly we find that, generally, the shortest merging timescales are obtained for those systems that predict fast rotating BHs. Indeed, systems in which the second-born BH is spun up or keeps a high rotation rate are those in which the tides are the strongest, which in turns translates to the shortest merging timescales.
From the discussion in sect. 2, the main contribution to $\chi_{\rm eff}$ (see Eq. 2) is from the spin $a_*$ of the second-born BH. $\chi_{\rm eff}$ thus decreases when the merging timescale increases. In other words, systems with small observed values of $\chi_{\rm eff}$ have a larger merging timescale, which can be seen
in Fig.~\ref{fig10}.
Finally, this trend implies that merging systems with a low $\chi_{\rm eff}$ are formed at high redshifts.

An anti-correlation between $\chi_{\rm eff}$ and the merging timescale was already predicted by the analytic models of \citet{2016MNRAS.462..844K} and \citet{2018MNRAS.473.4174Z}. However, our detailed calculations show that this anti-correlation is both more complex and weaker, as the relation of merging timescale to $\chi_{\rm eff}$ is also a function of the masses of the two binary components and the metallicity of the He-rich stars. It is interesting to see, in Fig.~\ref{fig10},  how $\chi_{\rm eff}$ varies with the chirp mass and the merging timescale. The chirp mass, M$_{\rm chirp}$, is defined by
    \begin{equation}\label{chirp}
    \centering
    M_{\rm chirp} = \frac{(m_{\rm 1} m_2)^{3/5}}{(m_{\rm 1}+m_2)^{1/5}},
    \end{equation}
where $m_1$ and $m_2$ are the masses of the two BHs, respectively.
Consistently with the discussion in Sect. 2, we have calculated $\chi_{\rm eff}$ assuming that the spin of the first-born BH is 0. A few interesting points can be noticed. First at solar metallicity, there is no possibility to produce chirp masses larger than about 17-18 M$_\odot$, even assuming that the first-born BH is $30\rm M_{\odot}$, which is unrealistic. In contrast, at low metallicities, provided the mass of the first born BH is high enough, there is no difficulty in producing chirp masses up to values around 30 M$_\odot$.

As already underlined above, high $\chi_{\rm eff}$ values are obtained only for short merging timescales. At low metallicities, it is easier to form BBHs with higher values of $\chi_{\rm eff}$. At the same time, these BBHs will likely have higher chirp masses, as the lower metallicity results in weaker wind mass-loss and larger overall final compact-object masses \citep[e.g.][]{2010ApJ...714.1217B}. However, low-metallicity star-formation environments are more common at high redshift \citep[e.g.][]{2014ApJ...791..130Z}. The combination of these correlations implies that BBHs with high  $\chi_{\rm eff}$ and M$_{\rm chirp}$ values have formed at high redshift (i.e. high $z_{\rm formation}$), but given their inferred short merging timescales, they have also merged at high redshift (i.e. high $z_{\rm observed}$). Given the current sensitivity of AdLIGO in the science runs O1 and O2, these merging BBHs are not detectable, as the highest observed redshift of a GW event is that of GW170104 at $z_{\rm observed}\simeq 0.18$. Future improvements in the sensitivity of GW observatories will allow the detection of GWs for BBH mergers at higher redshifts, and confirm or disprove the predicted complex correlation between  $\chi_{\rm eff}$, M$_{\rm chirp}$, merging timescale and metallicity, that is implied by the "CE" isolated binary formation channel.

\subsection{Comparisons with observed merging systems}

Five confirmed and one candidate GW events are believed to stem from the merger of two stellar mass BHs. In the ``CE'' channel, combining our results with the current six events, we draw the following conclusions:
\begin{itemize}
\item{The masses of the two BHs for GW150914 are around 30 M$_\odot$. Such ``heavy'' BHs are expected to form in metal-poor environment. We see that models with 0.1 $Z_\odot$ may provide a good match with the observed properties, at least as good as the one at $Z=0.01Z_\odot$. Such metallicities are encountered in the present day Universe within the Small Magellanic Cloud. Thus per se, a low metallicity does not strictly require a high redshift. On the other hand, low $\chi_{\rm eff}$ values imply a long merging timescale and thus imply that the merging occurred at high redshift. This illustrates how the information on the masses and the spin complement each other for constraining the metallicity and the redshift. From the events GW170104 and GW170814, although less extreme in term of BH masses, similar conclusions can be drawn.}
\item{The event LVT151012 has a lower chirp mass, nearly allowing solar metallicity models with a 30 M$_\odot$ BH to be compatible with its observed properties. However, even if it were possible to form a 30 M$_\odot$ BH at solar metallicity, present models predict  a merging timescale longer than the Hubble time. The measurement of $\chi_{\rm eff}$ consistent with zero allows for long merging timescales and hence, again, the formation of this BBH at high redshift seems to be the most likely scenario.}
\item{The event GW151226 may be explained at all metallicities considered here and in particular by solar metallicity models. Furthermore, taking at face value the statistically significant positive value for $\chi_{\rm eff}$, our models favor merger times shorter than a few Gyr. This implies that the system was formed at a redshift when most of the star formation occurs at solar-like metallicity.}
\item {GW170608 is the lowest-mass BBH merger yet reported. It is found that $\chi_{\rm eff}$ has a slight preference to be positive. Hence, the same arguments as for GW151226 hold.}
\end{itemize}

\section{Discussion}

The present results show that the spin parameter of the second-born BH can span all values between 0 and 1. Especially at lower metallicities, the dynamic range of initial orbital periods that lead to a final spin for the second-born BH between 0 and 1 is quite large. This result is different from those obtained by \citep{2018MNRAS.473.4174Z,2017ApJ...842..111H} who concluded that
the spin parameter of the second born BH will be either very low around 0, or very high around 1 (bimodality). The differences between the present results and those of the aforementioned studies stem from different assumptions concerning mass losses and tides. The bimodality results are based on the approach explained in \citet{2016MNRAS.462..844K}. Compared to the present approach, the impact of mass loss on $a_*$ is much less pronounced, and the impact of tides is much stronger. This is why, in their model, when tides are important, the He-rich star is tidally locked and the maximum spin is always reached. In our model, even if the orbit is tight at the beginning and tides are important, the star often cannot remain tidally locked indefinitely.

A main uncertainty of the ``CE'' isolated binary formation channel is attached to the way the CE phase is accounted for. In the present work, we did not study that phase directly since we began our investigation after the CE phase. We however implicitly assumed that
the star has lost its complete H-rich envelope and that the system is tight. These are
features commonly assumed as resulting from a CE phase and in that respect this work
follows the present common wisdom. Of course, would these facts be challenged by future studies, it would imply a very strong revision of the global scenario for the evolution of isolated close binary systems.

Mass loss due to the stellar winds of the He-rich star is another source of uncertainty.
We used here the most recent estimates for these mass loss rates and we did not explore
impact of changing these values. However, since we studied
the evolution of systems at three metallicities, and since changing the metallicity has
a deep impact on the mass loss rates, comparing the results at a range of metallicities gives an idea of what would be obtained by changing the mass loss rates. As discussed above, the effects are large and this underlines the fact that accurate mass loss rates are indeed needed to obtain reliable stellar evolution models.

The physics of the angular momentum transport inside stars is still uncertain. The present results have been obtained assuming a strong coupling between the core and the envelope mediated by a strong magnetic field \citep{1999A&A...349..189S,2002A&A...381..923S,2005ApJ...626..350H}. In case angular momentum would be transported mainly by meridional currents, the coupling is less efficient \citep[e.g.][]{2012A&A...542A..29G} allowing the core to retain more angular momentum. This would produce, all the other physical ingredients kept the same, faster rotating BHs and NSs.

\section{Conclusions}

Since the first GW event, GW150914, was discovered by AdLIGO, research in the field of BBH formation channels has been very active. The ``CE'' channel is one of the main proposed formation channels and likely the most widely studied one. The aim of this work is to investigate the final phase in the formation of BBH through this channel, namely the evolution of close binaries consisting of a He-rich star and a compact object. In doing so, we employed detailed binary evolution models that self-consistently take into account the effects of tidal interactions, wind mass loss and the evolution in the structure of the He star, including stellar rotation, and we explored a multidimensional parameter space. Our main findings can be summarized in the following:
\begin{itemize}

\item{Based on detailed stellar structure information, we computed the tidal coefficient E$_2$ for both H-rich and He-rich stars in a large range of masses and at three different metallicities (Z$_{\odot}$, 0.1Z${_\odot}$ and 0.01Z${_\odot}$). Based on those calculations, we derived fitting formulae that relate the value of E$_2$ to the ratio of the convective core radius to the total radius of the star.}

\item{We estimate that the spin of the first-born BH should be low ($a_{*,1}\lesssim 0.1$), as the progenitor star of the first-born BH evolves to the giant phase before loosing its envelope and collapsing to form a BH. During this expansion phase, most of the primordial angular momentum that the star might have had is transferred to its outer layers and subsequently lost due to Roche-lobe overflow mass transfer and wind mass loss. Hence, the first-born BH is expected to have a negligible contribution to the $\chi_{\rm eff}$ of the BBH, which in turns sets an upper limit to its possible value at $\chi_{\rm eff}\lesssim 0.5$}.

\item{The tidal synchronization timescale becomes comparable to or shorter than the lifetime of the He star at orbital periods below about $\sim$2 days, with the exact value depending on the mass of the He star and the binary mass ratio, but independent of metallicity. However, wind mass-losses are strongly dependent on metallicity, and hence the overall orbital evolution of the binary, which is determined by the interplay between tides and wind mass loss, does depend on metallicity.}

\item{Although we find that the initial rotation of the He star does affect our estimates for the resultant BH spin, arguments similar to the ones presented for the first-born BH imply that the expected initial rotation of the He star should be small ($\omega_{\rm init}\simeq 0-\omega_{\rm init,orb}$). Limiting the possible initial rotation of the He star in this range results in a weak dependence of our findings on that parameter.}

\item{A systematic exploration of the initial parameter space shows that the spin $a_*$ of the second-born BH covers the whole range (0-1, i.e from non-rotating to maximally rotating), and that especially at lower metallicities, the dynamical range of initial orbital periods that lead to BH spin with non-extreme values, i.e. between 0 and 1, is quite large.}

\item{Furthermore, we find an anti-correlation between the merging timescale of the BBH, T$_{\rm merger}$, and the spin of the second born BH, $a_*$, or the observable quantity $\chi_{\rm eff}$. This is a natural consequence of the fact that in order to form a fast rotating second-born BH, tides should be strong, and therefore the orbital separation between the He-star (progenitor of the second-born BH) and its compact-object companion should be small. The latter also leads to short merging timescales of the resulting BBH. We should note, however, that this anti-correlation is not a one-to-one relation between T$_{\rm merger}$ and $\chi_{\rm eff}$, as it also depends on other factors such as the masses of the two BHs, or the chirp mass M$_{\rm chirp}$ of the BBH, and the metallicity of the BH progenitor star. In that sense, simultaneous and precise estimates of M$_{\rm chirm}$, $\chi_{\rm eff}$ and the redshift at which the merger happened carry information about the time and environment at which the BBH was formed.}

\item{Our models present many possibilities for the formation of BBHs with non-zero, positive $\chi_{\rm eff}$. This at first glance is in contrast to the currently observed sample where 5 out the 6 detected merging BBHs have $\chi_{\rm eff}$ consistent with 0. However, one should also take into account the current sensitivity of GW observatories that limit us to mergers that happened in the local universe ($z_{\rm observed}\sim 0$) and show a strong preference to high chirp masses. Combining this with the star-formation and metallicity evolution of the Universe as a function of redshift, we conclude that, most likely, the currently observed sample of merging BBH mainly originates from BBHs that formed at low-metallicity environments and  $z_{\rm formation} \gtrsim 2-3$. Hence, these BBHs must have had long merging timescales and thus low $\chi_{\rm eff}$. As the sensitivity of GW observatories improves and we are able to probe more BBH mergers at high $z_{\rm observed}$ and/or lower chirp masses, our models predict that those BBHs will have preferentially positive, non-zero $\chi_{\rm eff}$}.

\end{itemize}

\begin{acknowledgements}
This work is sponsored by China Scholarship Council (CSC) and the Swiss National Science Foundation (project number 200020-172505). This project has received funding from the European Union’s Horizon 2020 research and innovation program under the Marie Sklodowska-Curie RISE action, grant agreement No 691164 (ASTROSTAT). TF acknowledges support from the Ambizione Fellowship of the Swiss National Science Foundation (grant PZ00P2\_ 148123). We  at DARK are grateful for support from the DNRF (Niels Bohr Professorship Program), the Carlsberg Foundation  and  the VILLUM FONDEN (project number 16599). JA acknowledges funding from the European Research Council under the European Union's Seventh Framework Programme (FP/2007-2013)/ERC Grant Agreement n. 617001. The computations were performed at the University of Geneva on the Baobab computer cluster. All figures were made with the free Python module Matplotlib \citep{2007CSE.....9...90H}.
\end{acknowledgements}

\bibliography{ref}

\begin{thebibliography}{96}
\expandafter\ifx\csname natexlab\endcsname\relax\def\natexlab#1{#1}\fi

\bibitem[{{Abbott} {et~al.}(2016{\natexlab{a}}){Abbott}, {Abbott}, {Abbott},
  {Abernathy}, {Acernese}, {Ackley}, {Adams}, {Adams}, {Addesso}, {Adhikari},
  \& et~al.}]{2016ApJ...818L..22A}
{Abbott}, B.~P., {Abbott}, R., {Abbott}, T.~D., {et~al.} 2016{\natexlab{a}},
  \apjl, 818, L22

\bibitem[{{Abbott} {et~al.}(2016{\natexlab{b}}){Abbott}, {Abbott}, {Abbott},
  {Abernathy}, {Acernese}, {Ackley}, {Adams}, {Adams}, {Addesso}, {Adhikari},
  \& et~al.}]{2016PhRvX...6d1015A}
{Abbott}, B.~P., {Abbott}, R., {Abbott}, T.~D., {et~al.} 2016{\natexlab{b}},
  Physical Review X, 6, 041015

\bibitem[{{Abbott} {et~al.}(2016{\natexlab{c}}){Abbott}, {Abbott}, {Abbott},
  {Abernathy}, {Acernese}, {Ackley}, {Adams}, {Adams}, {Addesso}, {Adhikari},
  \& et~al.}]{2016PhRvL.116x1103A}
{Abbott}, B.~P., {Abbott}, R., {Abbott}, T.~D., {et~al.} 2016{\natexlab{c}},
  Physical Review Letters, 116, 241103

\bibitem[{{Abbott} {et~al.}(2016{\natexlab{d}}){Abbott}, {Abbott}, {Abbott},
  {Abernathy}, {Acernese}, {Ackley}, {Adams}, {Adams}, {Addesso}, {Adhikari},
  \& et~al.}]{2016PhRvL.116f1102A}
{Abbott}, B.~P., {Abbott}, R., {Abbott}, T.~D., {et~al.} 2016{\natexlab{d}},
  Physical Review Letters, 116, 061102

\bibitem[{{Abbott} {et~al.}(2017{\natexlab{a}}){Abbott}, {Abbott}, {Abbott},
  {Acernese}, {Ackley}, {Adams}, {Adams}, {Addesso}, {Adhikari}, {Adya}, \&
  et~al.}]{2017PhRvL.118v1101A}
{Abbott}, B.~P., {Abbott}, R., {Abbott}, T.~D., {et~al.} 2017{\natexlab{a}},
  Physical Review Letters, 118, 221101

\bibitem[{{Abbott} {et~al.}(2017{\natexlab{b}}){Abbott}, {Abbott}, {Abbott},
  {Acernese}, {Ackley}, {Adams}, {Adams}, {Addesso}, {Adhikari}, {Adya}, \&
  et~al.}]{2017ApJ...851L..35A}
{Abbott}, B.~P., {Abbott}, R., {Abbott}, T.~D., {et~al.} 2017{\natexlab{b}},
  \apjl, 851, L35

\bibitem[{{Abbott} {et~al.}(2017{\natexlab{c}}){Abbott}, {Abbott}, {Abbott},
  {Acernese}, {Ackley}, {Adams}, {Adams}, {Addesso}, {Adhikari}, {Adya}, \&
  et~al.}]{2017PhRvL.119n1101A}
{Abbott}, B.~P., {Abbott}, R., {Abbott}, T.~D., {et~al.} 2017{\natexlab{c}},
  Physical Review Letters, 119, 141101

\bibitem[{{Abbott} {et~al.}(2017{\natexlab{d}}){Abbott}, {Abbott}, {Abbott},
  {Acernese}, {Ackley}, {Adams}, {Adams}, {Addesso}, {Adhikari}, {Adya}, \&
  et~al.}]{2017PhRvL.119p1101A}
{Abbott}, B.~P., {Abbott}, R., {Abbott}, T.~D., {et~al.} 2017{\natexlab{d}},
  Physical Review Letters, 119, 161101

\bibitem[{{Ackermann} {et~al.}(2016){Ackermann}, {Ajello}, {Albert},
  {Anderson}, {Arimoto}, {Atwood}, {Axelsson}, {Baldini}, {Ballet},
  {Barbiellini}, {Baring}, {Bastieri}, {Becerra Gonzalez}, {Bellazzini},
  {Bissaldi}, {Blandford}, {Bloom}, {Bonino}, {Bottacini}, {Brandt}, {Bregeon},
  {Britto}, {Bruel}, {Buehler}, {Burnett}, {Buson}, {Caliandro}, {Cameron},
  {Caputo}, {Caragiulo}, {Caraveo}, {Casandjian}, {Cavazzuti}, {Charles},
  {Chekhtman}, {Chiang}, {Chiaro}, {Ciprini}, {Cohen-Tanugi}, {Cominsky},
  {Condon}, {Costanza}, {Cuoco}, {Cutini}, {D'Ammando}, {de Palma}, {Desiante},
  {Digel}, {Di Lalla}, {Di Mauro}, {Di Venere}, {Dom{\'{\i}}nguez}, {Drell},
  {Dubois}, {Dumora}, {Favuzzi}, {Fegan}, {Ferrara}, {Franckowiak}, {Fukazawa},
  {Funk}, {Fusco}, {Gargano}, {Gasparrini}, {Gehrels}, {Giglietto}, {Giomi},
  {Giommi}, {Giordano}, {Giroletti}, {Glanzman}, {Godfrey}, {Gomez-Vargas},
  {Granot}, {Green}, {Grenier}, {Grondin}, {Grove}, {Guillemot}, {Guiriec},
  {Hadasch}, {Harding}, {Hays}, {Hewitt}, {Hill}, {Horan}, {Jogler},
  {J{\'o}hannesson}, {Kamae}, {Kensei}, {Kocevski}, {Kuss}, {La Mura},
  {Larsson}, {Latronico}, {Lemoine-Goumard}, {Li}, {Li}, {Longo}, {Loparco},
  {Lovellette}, {Lubrano}, {Madejski}, {Magill}, {Maldera}, {Manfreda},
  {Marelli}, {Mayer}, {Mazziotta}, {McEnery}, {Meyer}, {Michelson}, {Mirabal},
  {Mizuno}, {Moiseev}, {Monzani}, {Moretti}, {Morselli}, {Moskalenko},
  {Murgia}, {Negro}, {Nuss}, {Ohsugi}, {Omodei}, {Orienti}, {Orlando}, {Ormes},
  {Paneque}, {Perkins}, {Pesce-Rollins}, {Piron}, {Pivato}, {Porter},
  {Racusin}, {Rain{\`o}}, {Rando}, {Razzaque}, {Reimer}, {Reimer}, {Reposeur},
  {Ritz}, {Rochester}, {Romani}, {Saz Parkinson}, {Sgr{\`o}}, {Simone},
  {Siskind}, {Smith}, {Spada}, {Spandre}, {Spinelli}, {Suson}, {Tajima},
  {Thayer}, {Thayer}, {Thompson}, {Tibaldo}, {Torres}, {Troja}, {Uchiyama},
  {Venters}, {Vianello}, {Wood}, {Wood}, {Zaharijas}, {Zhu}, \&
  {Zimmer}}]{2016ApJ...823L...2A}
{Ackermann}, M., {Ajello}, M., {Albert}, A., {et~al.} 2016, \apjl, 823, L2

\bibitem[{{Bardeen} {et~al.}(1972){Bardeen}, {Press}, \&
  {Teukolsky}}]{1972ApJ...178..347B}
{Bardeen}, J.~M., {Press}, W.~H., \& {Teukolsky}, S.~A. 1972, \apj, 178, 347

\bibitem[{{Belczynski} {et~al.}(2010){Belczynski}, {Bulik}, {Fryer}, {Ruiter},
  {Valsecchi}, {Vink}, \& {Hurley}}]{2010ApJ...714.1217B}
{Belczynski}, K., {Bulik}, T., {Fryer}, C.~L., {et~al.} 2010, \apj, 714, 1217

\bibitem[{{Belczynski} {et~al.}(2016){Belczynski}, {Holz}, {Bulik}, \&
  {O'Shaughnessy}}]{2016Natur.534..512B}
{Belczynski}, K., {Holz}, D.~E., {Bulik}, T., \& {O'Shaughnessy}, R. 2016,
  \nat, 534, 512

\bibitem[{{Belczynski} {et~al.}(2017){Belczynski}, {Klencki}, {Meynet},
  {Fryer}, {Brown}, {Chruslinska}, {Gladysz}, {O'Shaughnessy}, {Bulik},
  {Berti}, {Holz}, {Gerosa}, {Giersz}, {Ekstrom}, {Georgy}, {Askar}, {Lasota},
  \& {Wysocki}}]{2017arXiv170607053B}
{Belczynski}, K., {Klencki}, J., {Meynet}, G., {et~al.} 2017, ArXiv e-prints

\bibitem[{{Brown} {et~al.}(2000){Brown}, {Lee}, {Wijers}, {Lee}, {Israelian},
  \& {Bethe}}]{2000NewA....5..191B}
{Brown}, G.~E., {Lee}, C.-H., {Wijers}, R.~A.~M.~J., {et~al.} 2000, \na, 5, 191

\bibitem[{{Cantiello} {et~al.}(2007){Cantiello}, {Yoon}, {Langer}, \&
  {Livio}}]{2007A&A...465L..29C}
{Cantiello}, M., {Yoon}, S.-C., {Langer}, N., \& {Livio}, M. 2007, \aap, 465,
  L29

\bibitem[{{Casares} \& {Jonker}(2014)}]{2014SSRv..183..223C}
{Casares}, J. \& {Jonker}, P.~G. 2014, \ssr, 183, 223

\bibitem[{{Claret}(2004)}]{2004A&A...424..919C}
{Claret}, A. 2004, \aap, 424, 919

\bibitem[{{Claret} \& {Cunha}(1997)}]{1997A&A...318..187C}
{Claret}, A. \& {Cunha}, N.~C.~S. 1997, \aap, 318, 187

\bibitem[{{Connaughton} {et~al.}(2016){Connaughton}, {Burns}, {Goldstein},
  {Blackburn}, {Briggs}, {Zhang}, {Camp}, {Christensen}, {Hui}, {Jenke},
  {Littenberg}, {McEnery}, {Racusin}, {Shawhan}, {Singer}, {Veitch},
  {Wilson-Hodge}, {Bhat}, {Bissaldi}, {Cleveland}, {Fitzpatrick}, {Giles},
  {Gibby}, {von Kienlin}, {Kippen}, {McBreen}, {Mailyan}, {Meegan}, {Paciesas},
  {Preece}, {Roberts}, {Sparke}, {Stanbro}, {Toelge}, \&
  {Veres}}]{2016ApJ...826L...6C}
{Connaughton}, V., {Burns}, E., {Goldstein}, A., {et~al.} 2016, \apjl, 826, L6

\bibitem[{{Darwin}(1879)}]{1879RSPS...29..168D}
{Darwin}, G.~H. 1879, Proceedings of the Royal Society of London Series I, 29,
  168

\bibitem[{{de Mink} {et~al.}(2009){de Mink}, {Cantiello}, {Langer}, {Pols},
  {Brott}, \& {Yoon}}]{2009A&A...497..243D}
{de Mink}, S.~E., {Cantiello}, M., {Langer}, N., {et~al.} 2009, \aap, 497, 243

\bibitem[{{de Mink} \& {Mandel}(2016)}]{2016MNRAS.460.3545D}
{de Mink}, S.~E. \& {Mandel}, I. 2016, \mnras, 460, 3545

\bibitem[{{Detmers} {et~al.}(2008){Detmers}, {Langer}, {Podsiadlowski}, \&
  {Izzard}}]{2008A&A...484..831D}
{Detmers}, R.~G., {Langer}, N., {Podsiadlowski}, P., \& {Izzard}, R.~G. 2008,
  \aap, 484, 831

\bibitem[{{D'Orazio} \& {Loeb}(2017)}]{2017arXiv170604211D}
{D'Orazio}, D.~J. \& {Loeb}, A. 2017, ArXiv e-prints

\bibitem[{{Eggleton}(1983)}]{1983ApJ...268..368E}
{Eggleton}, P.~P. 1983, \apj, 268, 368

\bibitem[{{Eldridge} {et~al.}(2008){Eldridge}, {Izzard}, \&
  {Tout}}]{2008MNRAS.384.1109E}
{Eldridge}, J.~J., {Izzard}, R.~G., \& {Tout}, C.~A. 2008, \mnras, 384, 1109

\bibitem[{{Eldridge} \& {Vink}(2006)}]{2006A&A...452..295E}
{Eldridge}, J.~J. \& {Vink}, J.~S. 2006, \aap, 452, 295

\bibitem[{{Farr} {et~al.}(2017){Farr}, {Stevenson}, {Miller}, {Mandel}, {Farr},
  \& {Vecchio}}]{2017Natur.548..426F}
{Farr}, W.~M., {Stevenson}, S., {Miller}, M.~C., {et~al.} 2017, \nat, 548, 426

\bibitem[{{Filippenko}(1997)}]{1997ARA&A..35..309F}
{Filippenko}, A.~V. 1997, \araa, 35, 309

\bibitem[{{Fragos} \& {McClintock}(2015)}]{2015ApJ...800...17F}
{Fragos}, T. \& {McClintock}, J.~E. 2015, \apj, 800, 17

\bibitem[{{Fryer}(1999)}]{1999ApJ...522..413F}
{Fryer}, C.~L. 1999, \apj, 522, 413

\bibitem[{{Gabriel} {et~al.}(2014){Gabriel}, {Noels}, {Montalb{\'a}n}, \&
  {Miglio}}]{2014A&A...569A..63G}
{Gabriel}, M., {Noels}, A., {Montalb{\'a}n}, J., \& {Miglio}, A. 2014, \aap,
  569, A63

\bibitem[{{Georgy} {et~al.}(2012){Georgy}, {Ekstr{\"o}m}, {Meynet}, {Massey},
  {Levesque}, {Hirschi}, {Eggenberger}, \& {Maeder}}]{2012A&A...542A..29G}
{Georgy}, C., {Ekstr{\"o}m}, S., {Meynet}, G., {et~al.} 2012, \aap, 542, A29

\bibitem[{{Heger} {et~al.}(2003){Heger}, {Fryer}, {Woosley}, {Langer}, \&
  {Hartmann}}]{2003ApJ...591..288H}
{Heger}, A., {Fryer}, C.~L., {Woosley}, S.~E., {Langer}, N., \& {Hartmann},
  D.~H. 2003, \apj, 591, 288

\bibitem[{{Heger} \& {Langer}(1998)}]{1998A&A...334..210H}
{Heger}, A. \& {Langer}, N. 1998, \aap, 334, 210

\bibitem[{{Heger} {et~al.}(2005){Heger}, {Woosley}, \&
  {Spruit}}]{2005ApJ...626..350H}
{Heger}, A., {Woosley}, S.~E., \& {Spruit}, H.~C. 2005, \apj, 626, 350

\bibitem[{{Hotokezaka} \& {Piran}(2017{\natexlab{a}})}]{2017arXiv170708978H}
{Hotokezaka}, K. \& {Piran}, T. 2017{\natexlab{a}}, ArXiv e-prints

\bibitem[{{Hotokezaka} \& {Piran}(2017{\natexlab{b}})}]{2017ApJ...842..111H}
{Hotokezaka}, K. \& {Piran}, T. 2017{\natexlab{b}}, \apj, 842, 111

\bibitem[{{Hunter}(2007)}]{2007CSE.....9...90H}
{Hunter}, J.~D. 2007, Computing in Science and Engineering, 9, 90

\bibitem[{{Hurley} {et~al.}(2002){Hurley}, {Tout}, \&
  {Pols}}]{2002MNRAS.329..897H}
{Hurley}, J.~R., {Tout}, C.~A., \& {Pols}, O.~R. 2002, \mnras, 329, 897

\bibitem[{{Hut}(1981)}]{1981A&A....99..126H}
{Hut}, P. 1981, \aap, 99, 126

\bibitem[{{Inayoshi} {et~al.}(2017){Inayoshi}, {Hirai}, {Kinugawa}, \&
  {Hotokezaka}}]{2017MNRAS.468.5020I}
{Inayoshi}, K., {Hirai}, R., {Kinugawa}, T., \& {Hotokezaka}, K. 2017, \mnras,
  468, 5020

\bibitem[{{Izzard} {et~al.}(2004){Izzard}, {Ramirez-Ruiz}, \&
  {Tout}}]{2004MNRAS.348.1215I}
{Izzard}, R.~G., {Ramirez-Ruiz}, E., \& {Tout}, C.~A. 2004, \mnras, 348, 1215

\bibitem[{{Kocsis} \& {Levin}(2012)}]{2012PhRvD..85l3005K}
{Kocsis}, B. \& {Levin}, J. 2012, \prd, 85, 123005

\bibitem[{{Kushnir} {et~al.}(2016){Kushnir}, {Zaldarriaga}, {Kollmeier}, \&
  {Waldman}}]{2016MNRAS.462..844K}
{Kushnir}, D., {Zaldarriaga}, M., {Kollmeier}, J.~A., \& {Waldman}, R. 2016,
  \mnras, 462, 844

\bibitem[{{Kushnir} {et~al.}(2017){Kushnir}, {Zaldarriaga}, {Kollmeier}, \&
  {Waldman}}]{2017MNRAS.467.2146K}
{Kushnir}, D., {Zaldarriaga}, M., {Kollmeier}, J.~A., \& {Waldman}, R. 2017,
  \mnras, 467, 2146

\bibitem[{{Langer}(1997)}]{1997ASPC..120...83L}
{Langer}, N. 1997, in Astronomical Society of the Pacific Conference Series,
  Vol. 120, Luminous Blue Variables: Massive Stars in Transition, ed. A.~{Nota}
  \& H.~{Lamers}, 83

\bibitem[{{Langer}(1998)}]{1998A&A...329..551L}
{Langer}, N. 1998, \aap, 329, 551

\bibitem[{{LIGO Scientific Collaboration} {et~al.}(2015){LIGO Scientific
  Collaboration}, {Aasi}, {Abbott}, {Abbott}, {Abbott}, {Abernathy}, {Ackley},
  {Adams}, {Adams}, {Addesso}, \& et~al.}]{2015CQGra..32g4001L}
{LIGO Scientific Collaboration}, {Aasi}, J., {Abbott}, B.~P., {et~al.} 2015,
  Classical and Quantum Gravity, 32, 074001

\bibitem[{{Loeb}(2016)}]{2016ApJ...819L..21L}
{Loeb}, A. 2016, \apjl, 819, L21

\bibitem[{{Maeder} \& {Meynet}(2000)}]{2000A&A...361..159M}
{Maeder}, A. \& {Meynet}, G. 2000, \aap, 361, 159

\bibitem[{{Mandel} \& {de Mink}(2016)}]{2016MNRAS.458.2634M}
{Mandel}, I. \& {de Mink}, S.~E. 2016, \mnras, 458, 2634

\bibitem[{{Marchant} {et~al.}(2016){Marchant}, {Langer}, {Podsiadlowski},
  {Tauris}, \& {Moriya}}]{2016A&A...588A..50M}
{Marchant}, P., {Langer}, N., {Podsiadlowski}, P., {Tauris}, T.~M., \&
  {Moriya}, T.~J. 2016, \aap, 588, A50

\bibitem[{{McClintock}(2006)}]{2006AAS...208.3301M}
{McClintock}, J.~E. 2006, in Bulletin of the American Astronomical Society,
  Vol.~38, American Astronomical Society Meeting Abstracts \#208, 109

\bibitem[{{McClintock} {et~al.}(2014){McClintock}, {Narayan}, \&
  {Steiner}}]{2014SSRv..183..295M}
{McClintock}, J.~E., {Narayan}, R., \& {Steiner}, J.~F. 2014, \ssr, 183, 295

\bibitem[{{Miller} \& {Lauburg}(2009)}]{2009ApJ...692..917M}
{Miller}, M.~C. \& {Lauburg}, V.~M. 2009, \apj, 692, 917

\bibitem[{{Miller} \& {Miller}(2015)}]{2015PhR...548....1M}
{Miller}, M.~C. \& {Miller}, J.~M. 2015, \physrep, 548, 1

\bibitem[{{Novikov} \& {Thorne}(1973)}]{1973blho.conf..343N}
{Novikov}, I.~D. \& {Thorne}, K.~S. 1973, in Black Holes (Les Astres Occlus),
  ed. C.~{Dewitt} \& B.~S. {Dewitt}, 343--450

\bibitem[{{O'Leary} {et~al.}(2009){O'Leary}, {Kocsis}, \&
  {Loeb}}]{2009MNRAS.395.2127O}
{O'Leary}, R.~M., {Kocsis}, B., \& {Loeb}, A. 2009, \mnras, 395, 2127

\bibitem[{{Paxton} {et~al.}(2011){Paxton}, {Bildsten}, {Dotter}, {Herwig},
  {Lesaffre}, \& {Timmes}}]{2011ApJS..192....3P}
{Paxton}, B., {Bildsten}, L., {Dotter}, A., {et~al.} 2011, \apjs, 192, 3

\bibitem[{{Paxton} {et~al.}(2013){Paxton}, {Cantiello}, {Arras}, {Bildsten},
  {Brown}, {Dotter}, {Mankovich}, {Montgomery}, {Stello}, {Timmes}, \&
  {Townsend}}]{2013ApJS..208....4P}
{Paxton}, B., {Cantiello}, M., {Arras}, P., {et~al.} 2013, \apjs, 208, 4

\bibitem[{{Paxton} {et~al.}(2015){Paxton}, {Marchant}, {Schwab}, {Bauer},
  {Bildsten}, {Cantiello}, {Dessart}, {Farmer}, {Hu}, {Langer}, {Townsend},
  {Townsley}, \& {Timmes}}]{2015ApJS..220...15P}
{Paxton}, B., {Marchant}, P., {Schwab}, J., {et~al.} 2015, \apjs, 220, 15

\bibitem[{{Paxton} {et~al.}(2018){Paxton}, {Schwab}, {Bauer}, {Bildsten},
  {Blinnikov}, {Duffell}, {Farmer}, {Goldberg}, {Marchant}, {Sorokina},
  {Thoul}, {Townsend}, \& {Timmes}}]{2018ApJS..234...34P}
{Paxton}, B., {Schwab}, J., {Bauer}, E.~B., {et~al.} 2018, \apjs, 234, 34

\bibitem[{{Peters}(1964)}]{1964PhRv..136.1224P}
{Peters}, P.~C. 1964, Physical Review, 136, 1224

\bibitem[{{Petrovic} {et~al.}(2005){Petrovic}, {Langer}, \& {van der
  Hucht}}]{2005A&A...435.1013P}
{Petrovic}, J., {Langer}, N., \& {van der Hucht}, K.~A. 2005, \aap, 435, 1013

\bibitem[{{Petrovich} \& {Antonini}(2017)}]{2017ApJ...846..146P}
{Petrovich}, C. \& {Antonini}, F. 2017, \apj, 846, 146

\bibitem[{{Phinney}(1991)}]{1991ApJ...380L..17P}
{Phinney}, E.~S. 1991, \apjl, 380, L17

\bibitem[{{Planck Collaboration} {et~al.}(2016){Planck Collaboration}, {Ade},
  {Aghanim}, {Arnaud}, {Ashdown}, {Aumont}, {Baccigalupi}, {Banday},
  {Barreiro}, {Bartlett}, \& et~al.}]{2016A&A...594A..13P}
{Planck Collaboration}, {Ade}, P.~A.~R., {Aghanim}, N., {et~al.} 2016, \aap,
  594, A13

\bibitem[{{Rasio} {et~al.}(1996){Rasio}, {Tout}, {Lubow}, \&
  {Livio}}]{1996ApJ...470.1187R}
{Rasio}, F.~A., {Tout}, C.~A., {Lubow}, S.~H., \& {Livio}, M. 1996, \apj, 470,
  1187

\bibitem[{{Reynolds}(2014)}]{2014SSRv..183..277R}
{Reynolds}, C.~S. 2014, \ssr, 183, 277

\bibitem[{{Rodriguez} {et~al.}(2016){Rodriguez}, {Haster}, {Chatterjee},
  {Kalogera}, \& {Rasio}}]{2016ApJ...824L...8R}
{Rodriguez}, C.~L., {Haster}, C.-J., {Chatterjee}, S., {Kalogera}, V., \&
  {Rasio}, F.~A. 2016, \apjl, 824, L8

\bibitem[{{Rodriguez} {et~al.}(2015){Rodriguez}, {Morscher}, {Pattabiraman},
  {Chatterjee}, {Haster}, \& {Rasio}}]{2015PhRvL.115e1101R}
{Rodriguez}, C.~L., {Morscher}, M., {Pattabiraman}, B., {et~al.} 2015, Physical
  Review Letters, 115, 051101

\bibitem[{{Savchenko} {et~al.}(2016){Savchenko}, {Ferrigno}, {Mereghetti},
  {Natalucci}, {Bazzano}, {Bozzo}, {Brandt}, {Courvoisier}, {Diehl}, {Hanlon},
  {von Kienlin}, {Kuulkers}, {Laurent}, {Lebrun}, {Roques}, {Ubertini}, \&
  {Weidenspointner}}]{2016ApJ...820L..36S}
{Savchenko}, V., {Ferrigno}, C., {Mereghetti}, S., {et~al.} 2016, \apjl, 820,
  L36

\bibitem[{{Siess} {et~al.}(2013){Siess}, {Izzard}, {Davis}, \&
  {Deschamps}}]{2013A&A...550A.100S}
{Siess}, L., {Izzard}, R.~G., {Davis}, P.~J., \& {Deschamps}, R. 2013, \aap,
  550, A100

\bibitem[{{Sigurdsson} \& {Hernquist}(1993)}]{1993Natur.364..423S}
{Sigurdsson}, S. \& {Hernquist}, L. 1993, \nat, 364, 423

\bibitem[{{Song} {et~al.}(2013){Song}, {Maeder}, {Meynet}, {Huang},
  {Ekstr{\"o}m}, \& {Granada}}]{2013A&A...556A.100S}
{Song}, H.~F., {Maeder}, A., {Meynet}, G., {et~al.} 2013, \aap, 556, A100

\bibitem[{{Song} {et~al.}(2016){Song}, {Meynet}, {Maeder}, {Ekstr{\"o}m}, \&
  {Eggenberger}}]{2016A&A...585A.120S}
{Song}, H.~F., {Meynet}, G., {Maeder}, A., {Ekstr{\"o}m}, S., \& {Eggenberger},
  P. 2016, \aap, 585, A120

\bibitem[{{Song} {et~al.}(2018){Song}, {Meynet}, {Maeder}, {Ekstr{\"o}m},
  {Eggenberger}, {Georgy}, {Qin}, {Fragos}, {Soerensen}, {Barblan}, \&
  {Wade}}]{2018A&A...609A...3S}
{Song}, H.~F., {Meynet}, G., {Maeder}, A., {et~al.} 2018, \aap, 609, A3

\bibitem[{{Spruit}(1999)}]{1999A&A...349..189S}
{Spruit}, H.~C. 1999, \aap, 349, 189

\bibitem[{{Spruit}(2002)}]{2002A&A...381..923S}
{Spruit}, H.~C. 2002, \aap, 381, 923

\bibitem[{{Sukhbold} {et~al.}(2016){Sukhbold}, {Ertl}, {Woosley}, {Brown}, \&
  {Janka}}]{2016ApJ...821...38S}
{Sukhbold}, T., {Ertl}, T., {Woosley}, S.~E., {Brown}, J.~M., \& {Janka}, H.-T.
  2016, \apj, 821, 38

\bibitem[{{Toledano} {et~al.}(2007){Toledano}, {Moreno}, {Koenigsberger},
  {Detmers}, \& {Langer}}]{2007A&A...461.1057T}
{Toledano}, O., {Moreno}, E., {Koenigsberger}, G., {Detmers}, R., \& {Langer},
  N. 2007, \aap, 461, 1057

\bibitem[{{Tutukov} \& {Yungelson}(1973)}]{1973NInfo..27...70T}
{Tutukov}, A. \& {Yungelson}, L. 1973, Nauchnye Informatsii, 27, 70

\bibitem[{{Tutukov} \& {Yungelson}(1993)}]{1993MNRAS.260..675T}
{Tutukov}, A.~V. \& {Yungelson}, L.~R. 1993, \mnras, 260, 675

\bibitem[{{van den Heuvel} {et~al.}(2017){van den Heuvel}, {Portegies Zwart},
  \& {de Mink}}]{2017MNRAS.471.4256V}
{van den Heuvel}, E.~P.~J., {Portegies Zwart}, S.~F., \& {de Mink}, S.~E. 2017,
  \mnras, 471, 4256

\bibitem[{{van den Heuvel} \& {Yoon}(2007)}]{2007Ap&SS.311..177V}
{van den Heuvel}, E.~P.~J. \& {Yoon}, S.-C. 2007, \apss, 311, 177

\bibitem[{{Woosley}(1993)}]{1993AAS...182.5505W}
{Woosley}, S.~E. 1993, in Bulletin of the American Astronomical Society,
  Vol.~25, American Astronomical Society Meeting Abstracts \#182, 894

\bibitem[{{Woosley}(2016)}]{2016ApJ...824L..10W}
{Woosley}, S.~E. 2016, \apjl, 824, L10

\bibitem[{{Woosley} \& {Bloom}(2006)}]{2006ARA&A..44..507W}
{Woosley}, S.~E. \& {Bloom}, J.~S. 2006, \araa, 44, 507

\bibitem[{{Wu} {et~al.}(2013){Wu}, {Hou}, \& {Lei}}]{2013ApJ...767L..36W}
{Wu}, X.-F., {Hou}, S.-J., \& {Lei}, W.-H. 2013, \apjl, 767, L36

\bibitem[{{Yoon} {et~al.}(2006){Yoon}, {Langer}, \&
  {Norman}}]{2006A&A...460..199Y}
{Yoon}, S.-C., {Langer}, N., \& {Norman}, C. 2006, \aap, 460, 199

\bibitem[{{Yoon} {et~al.}(2010){Yoon}, {Woosley}, \&
  {Langer}}]{2010ApJ...725..940Y}
{Yoon}, S.-C., {Woosley}, S.~E., \& {Langer}, N. 2010, \apj, 725, 940

\bibitem[{{Zahid} {et~al.}(2014){Zahid}, {Dima}, {Kudritzki}, {Kewley},
  {Geller}, {Hwang}, {Silverman}, \& {Kashino}}]{2014ApJ...791..130Z}
{Zahid}, H.~J., {Dima}, G.~I., {Kudritzki}, R.-P., {et~al.} 2014, \apj, 791,
  130

\bibitem[{{Zahn}(1975)}]{1975A&A....41..329Z}
{Zahn}, J.-P. 1975, \aap, 41, 329

\bibitem[{{Zahn}(1977)}]{1977A&A....57..383Z}
{Zahn}, J.-P. 1977, \aap, 57, 383

\bibitem[{{Zaldarriaga} {et~al.}(2018){Zaldarriaga}, {Kushnir}, \&
  {Kollmeier}}]{2018MNRAS.473.4174Z}
{Zaldarriaga}, M., {Kushnir}, D., \& {Kollmeier}, J.~A. 2018, \mnras, 473, 4174

\end{thebibliography}

\begin{appendix}
\section{Tidal coefficient E$_n$}
Dynamical tides are dominant for stars with radiative envelopes and convective cores \citep{1977A&A....57..383Z}. The synchronization and circularization timescales depends on the tidal torque coefficient E$_n$ (only $n = 2$ is considered, as the contributions from larger $n$ are negligible). The complete set of equations to calculate $E_n$ are as follows \citep{1977A&A....57..383Z}:
  \begin{equation}\label{eqa1}
        E_n = \frac{3^{8/3}[\Gamma(4/3)]^2}{(2n+1)[n(n+1)]^{4/3}} \frac{\rho_fR^3}{M} \left[\frac{R}{g_s} \left(\frac{-gB}{x^2} \right)'_f\right]^{-1/3}H^2_n,
  \end{equation}
where $\Gamma$ is the usual gamma function, $f$ and $s$ refer to the boundary of the convective core and surface, respectively.  $x$ denotes the normalized radius coordinate of the star, i.e. $x = r/R$. Primed symbols denote derivatives with respect to $x$,
$R$ is the stellar radius, $M$ the stellar mass, $g$ the gravity, and $-gB$ is the square of the Brunt-V{\"a}is{\"a}l{\"a} frequency. $B$ is the difference between the actual density gradient and the adiabatic one:
  \begin{equation}\label{eqa2}
  \centering
        B = \frac{\mathrm{d}}{\mathrm{d} r} {\rm ln}\rho - \frac{1}{\Gamma_1}\frac{\mathrm{d}}{\mathrm{d} r} {\rm ln} P,
  \end{equation}
where $\Gamma_1$ is the adiabatic exponent $\left(\frac{d {\rm ln} P}{d {\rm ln} \rho}\right)_{ad}$.
The coefficient H$_n$ is given by
  \begin{equation}\label{eqa3}
  \centering
        H_n = \frac{1}{X(x_f)Y(1)}\int_{0}^{x_f}\left[Y'' -\frac{n(n+1)Y}{x^2}\right]X dx,
  \end{equation}
where $X$ and $Y$ are found by solving the following second order differential equations:
  \begin{equation}\label{eqa4}
  \centering
  \begin{split}
        X'' - \frac{\rho'}{\rho}X' - \frac{n(n+1)}{x^2}X=0, \\
        Y'' -6(1-\frac{\rho}{\bar{\rho}})\frac{Y'}{x}-\left[n(n+1)-12\left(1-\frac{\rho}{\bar{\rho}}\right)\right]\frac{Y}{x^2}=0,
  \end{split}
  \end{equation}
where $\bar{\rho}$ is the mean density inside radius $r$ of the star and the primes on $X$ and $Y$ indicate derivatives with respect to $x$. More details about the initial conditions for solving the differential equations above can be found in \citet{2013A&A...550A.100S}.

A fourth order adaptive stepsize Runge-Kutta method is used to solve the two differential equations for $X$ and $Y$. We have found that the derivative of the Brunt-V{\"a}is{\"a}l{\"a} frequency divided by x$^2$ in Eq. \ref{eqa1} is sensitive to the boundary of the convective core. For determining the boundary of the convective cores, which define $f$ and $s$ in Eq. \ref{eqa1}, we used the Schwarzschild criterion. Overshooting above the Schwarzschild boundary of the convective core is considered with an extension
given by $\alpha_p$ = 0.1 H$_p$, where H$_p$ is the pressure scale height estimated at the Schwarzschild boundary limit.
  \begin{figure}[h]
     \centering
     \includegraphics[width=\columnwidth]{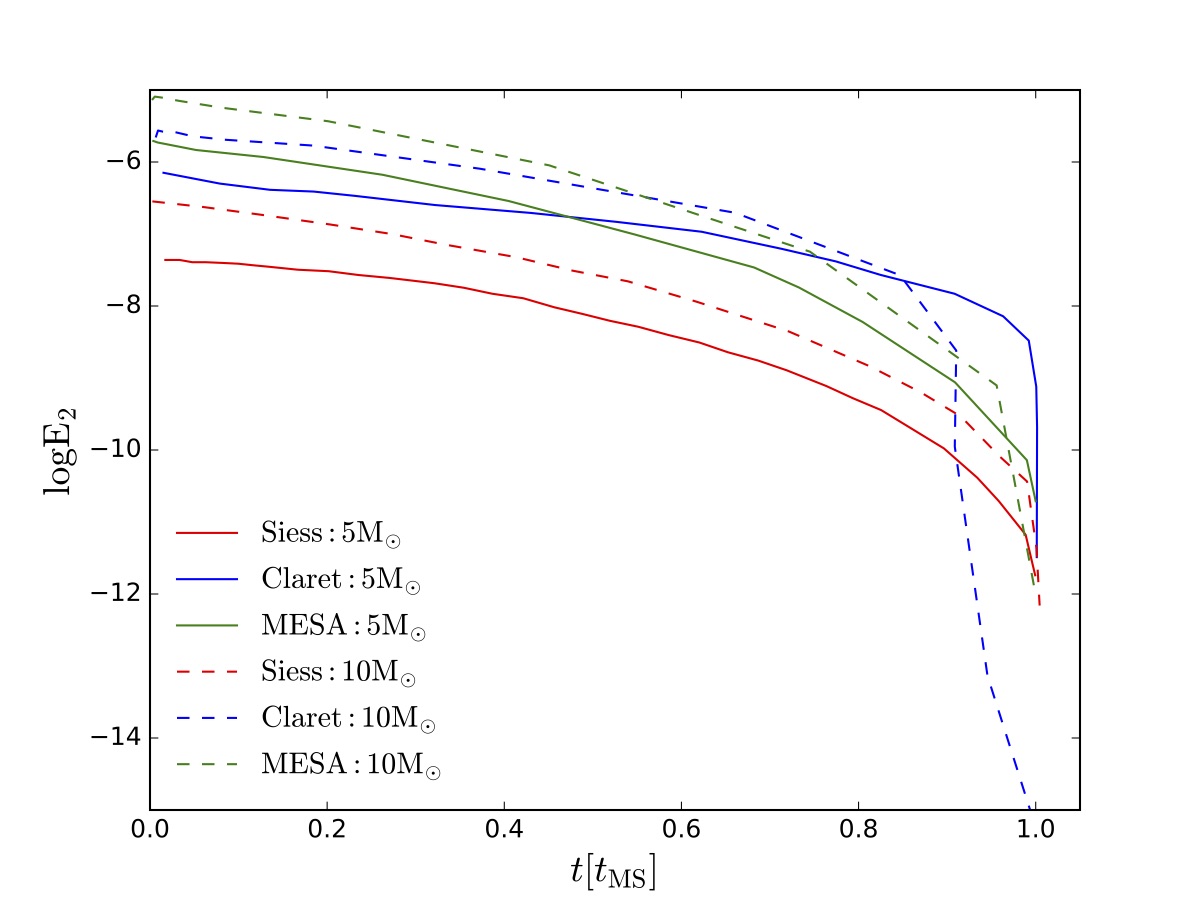}
     \caption{Comparison of E$_2$ coefficients computed by different authors. Solid lines and dashed lines correspond to 5 and 10 M$_\odot$ main-sequence stars, respectively. Red: Data from \citet{2013A&A...550A.100S}; blue: Data from \citet{2004A&A...424..919C}; green:  results from this study.}
     \label{figa1}
  \end{figure}

In Fig.~\ref{figa1} we show the evolution of E$_2$ for a 5 and 10 M$_\odot$ H-rich star at solar metallicity, as computed by the method described above. We compare our results with calculations by \citet{2013A&A...550A.100S} and \citet{1997A&A...318..187C}. As it was pointed out by \citet{1977A&A....57..383Z}, E$_2$ is sensitive to the exact structure of the star, something which was also confirmed by \citet{1997A&A...318..187C}. Different treatments for the boundary of the convective core
may lead to slight differences in the size of the convective core, but also, even when using the same physical criterion, the numerical implementation may differ \citep{2014A&A...569A..63G}. The adopted overshooting parameter $\alpha_{ov}$, which affects the extent of the convective core, may also be responsible for differences in the evolution of E$_2$. Compared with our value of 0.1H$_p$, \citet{2013A&A...550A.100S} used a value for $\alpha_{ov}$ inthe range of 0.23-0.3H$_p$, while \citet{1997A&A...318..187C} adopted a value of 0.2H$_p$.

\begin{figure*}[h]
     \centering
     \includegraphics[width=0.99\textwidth]{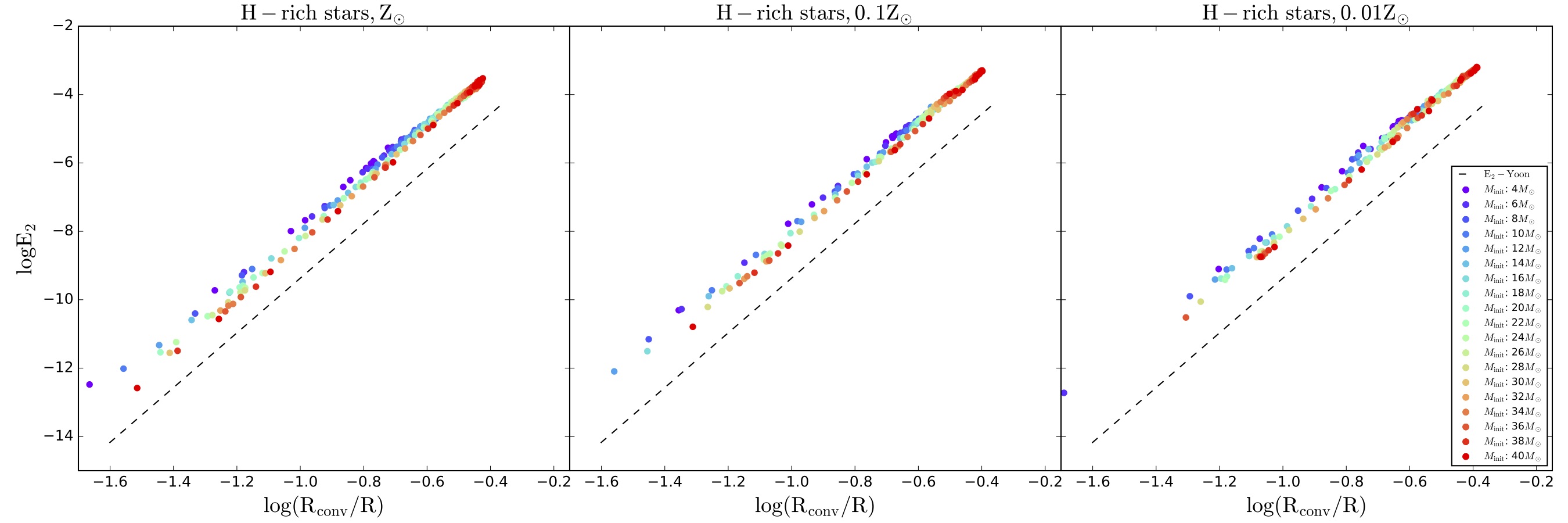}
     \caption{E$_2$ as a function of R$_{\rm conv}$/R for H-rich stars at different metallicities. The different colors of the points correspond to various masses of the stars. Black dotted line corresponds to the formula from \citet{2010ApJ...725..940Y}. Left panel: Z$_\odot$, middle panel: 0.1 Z$_\odot$, right panel: 0.01 Z$_\odot$.}
     \label{figa2}
  \end{figure*}

\begin{figure*}[h]
     \centering
     \includegraphics[width=0.99\textwidth]{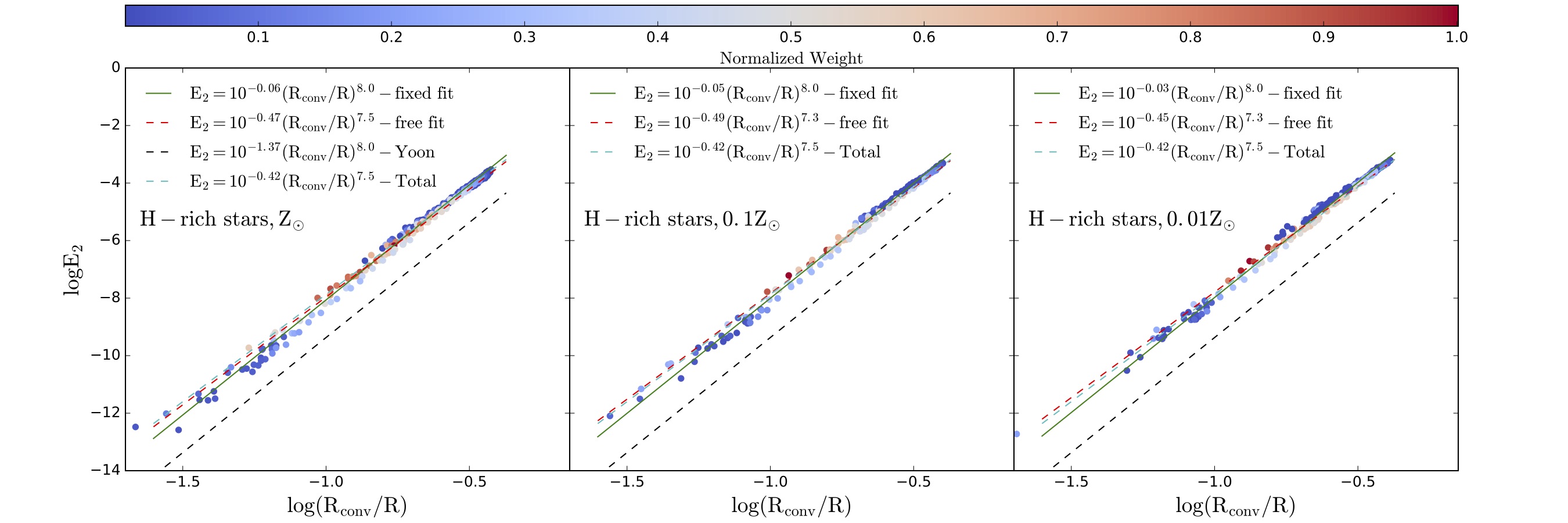}
     \caption{E$_2$ as a function of R$_{\rm conv}$/R for H-rich stars at different metallicities. Black dotted line corresponds to the formula from \citet{2010ApJ...725..940Y}; red dashed line refers to the free fitting; green solid line refers to the fitting data with the fixed exponent of  R$_{\rm conv}$/R = 8.0, cyan dotted line refers to the fitting data with all three different metallicities. Different color bar points correspond to the weights of E$_2$ defined in Eq. \ref{eqa5}. Left panel: Z$_\odot$, middle panel: 0.1 Z$_\odot$, right panel: 0.01 Z$_\odot$.}
     \label{figa3}
  \end{figure*}

For stars on the main sequence, E$_2$ depends strongly on the radius of the convective core R$_{\rm conv}$. In Fig.~\ref{figa2}, the relation between E$_2$ and the ratio of R$_{\rm conv}$ to the total radius of the star is shown for star with masses between 2 and 40 M$_\odot$ and for metallicity Z = 0.01 Z$_\odot$, 0.1 Z$_\odot$ and Z$_\odot$. Our estimates of E$_2$ are offset by about one magnitude above the relation given by \citet{2010ApJ...725..940Y}, but in closer agreement with calculations by \citet{1997A&A...318..187C}. Figure A.3 shows the fitting formulae that can be deduced from our calculations of H-rich stars. Note that for a given initial mass, some points correspond to long time-steps of our stellar evolution code and other points correspond to very short ones. To account for this effect when deriving our fitting formulae, we weight each of the data points by:
  \begin{equation}\label{eqa5}
      \centering
     weight = \frac{dt}{T} * \frac{1}{N},
  \end{equation}
where $dt$ is the time step, $N$ number of the steps and $T$ the lifetime of the star during the core H-burning phase. On each panel corresponding to one specific metallicity, we have used three fitting methods. First, in order to have a comparison with the result of Yoon et al., we fixed the exponent relating E$_2$ with R$_{\rm conv}$ to 8, and performed a separate fit for each metallicity. This result is shown by the green solid line in Fig.~\ref{figa3}. Second, we directly fitted the data allowing for the exponent to vary freely. This is shown as a red dashed line in Fig.~\ref{figa3}. Third, we fitted the combined data for all metallicities together. This is shown as a cyan dashed line. We find  only small differences between the three fitting methods and a negligible dependence on the metallicity. Therefore the same fitting formula is suggested for use across metallicities. For reference, we also provide Yoon et al.'s fitting formula as a black, dashed line.

\begin{figure*}[h]
     \centering
     \includegraphics[width=0.99\textwidth]{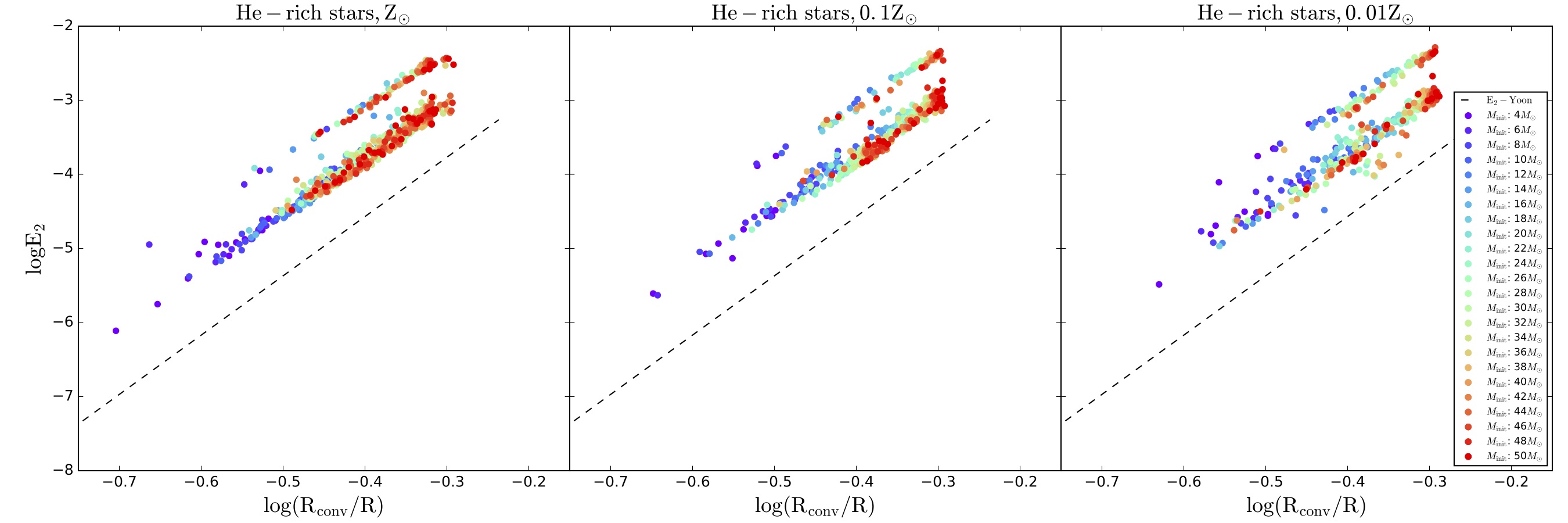}
     \caption{Same as Fig.~\ref{figa2} for He-rich stars.}
     \label{figa4}
  \end{figure*}

\begin{figure*}[h]
     \centering
     \includegraphics[width=0.99\textwidth]{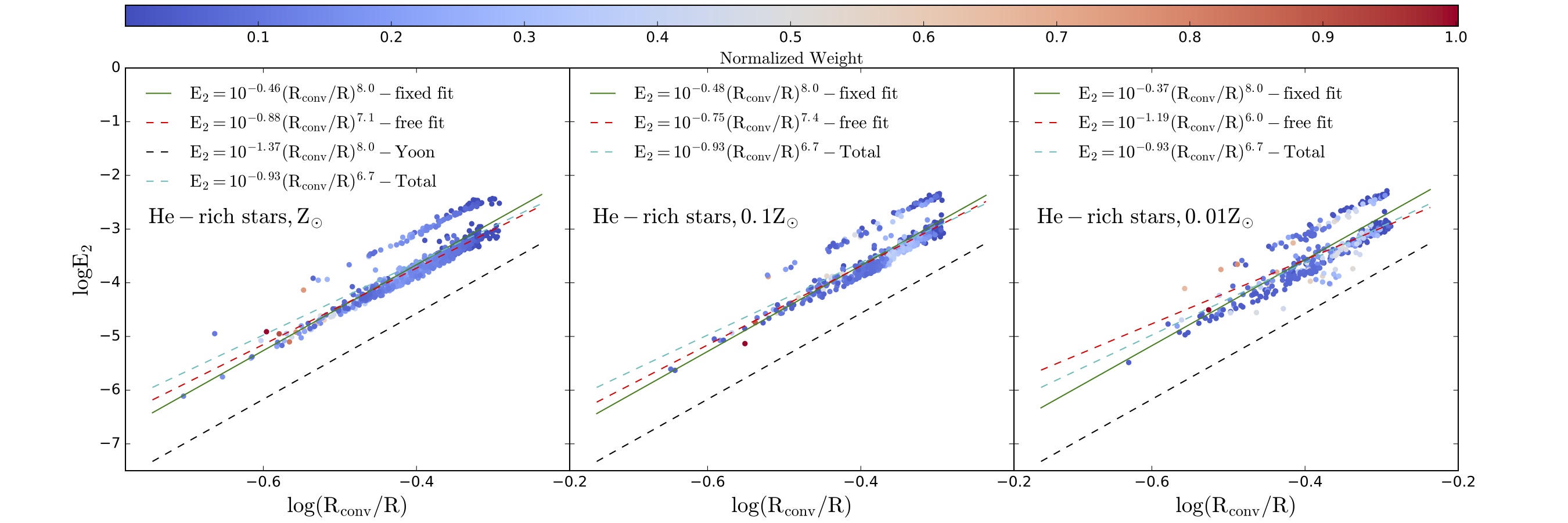}
     \caption{Same as Fig.~\ref{figa3} for He-rich stars.}
     \label{figa5}
  \end{figure*}

We have also investigated the relation of E$_2$ and R$_{\rm conv}/$R for He-rich stars at different metallicities. The masses of He-rich stars in our investigation cover the range from 4 to 50 M$_\odot$ with a mass interval of 2 M$_\odot$. Similarly, the results for He-rich stars are shown in Figs.~\ref{figa4} and \ref{figa5}. On these figures, we can see some ``jumps'' in the calculated values of E$_2$ which are due to the unstable boundary of the convective core. However, the fitting results are not significantly influenced by these jumps when the weights of the data points are considered. Given the insensitivity of our calculations to the metallicity, we again suggest that the same fitting formula should be used for all He-rich stars irrespective of their metallicity.

Ideally, E$_2$ should be calculated at every time step since its value depends on the structure of the star which evolves as a function of time and also depends on the important physical ingredients that vary from one set of models to another. Hence, it is not advisable to use a published formula without at least checking the conditions that have been used to obtain it.

Here we have established new fitting formulae that correspond to the physics of the present stellar models. We have shown that our fitting formulae differ significantly from the one proposed by \citet{2010ApJ...725..940Y}. The difference comes mainly from the treatment of convective criterion,  the boundary of the convective core, as well as the overshooting. We show that the fitting formula for the H- and He-rich stars are somewhat different; however these formulae do not strongly depend on the metallicity of the star.

  \end{appendix}

\end{document}